\documentclass[preprint2]{aastex631}

\usepackage{graphicx}
\graphicspath {./figs/}
\usepackage{times}
\usepackage[utf8]{inputenc}
\usepackage{enumitem,amssymb}
\usepackage{tcolorbox}

\usepackage[version=4]{mhchem} 
\usepackage[nameinlink]{cleveref}
\usepackage[para,online,flushleft]{threeparttable}
\usepackage{xspace}
\usepackage{upgreek}
\newcommand{\um}[0]{$\upmu$m\xspace}

\newcommand{\smarter}{\texttt{smarter}\xspace}
\newcommand{\smart}{\texttt{smart}\xspace}
\newcommand{\lblabc}{\texttt{lblabc}\xspace}
\newcommand{\revision}[1]{{#1}}
\defcitealias{Stevenson2020}{S20}
\defcitealias{Lustig-Yaeger2021}{LY21}

\shorttitle{MIRECLE: Retrievals and Science Yield}
\shortauthors{Mandell et al.}

\begin{document}


\title{MIRECLE: Science Yield for a Mid-IR Explorer-Class Mission to Study Non-Transiting Rocky Planets Orbiting the Nearest M-Stars Using Planetary Infrared Excess}

\author[0000-0002-0786-7307]{Avi M. Mandell}
\affiliation{NASA Goddard Space Flight Center, 8800 Greenbelt Rd, Greenbelt, MD 20771, USA}
\affiliation{GSFC Sellers Exoplanet Environments Collaboration}

\author[0000-0002-0746-1980]{Jacob Lustig-Yaeger}
\affiliation{Johns Hopkins University Applied Physics Laboratory, Laurel, MD 20723, USA}
\affiliation{NASA NExSS Virtual Planetary Laboratory, Box 351580, University of Washington, Seattle, Washington 98195, USA}

\author[0000-0002-7352-7941]{Kevin B. Stevenson}
\affiliation{Johns Hopkins University Applied Physics Laboratory, Laurel, MD 20723, USA}
\affiliation{Johns Hopkins University, 3400 N. Charles St, Baltimore, MD 21218, USA}

\author[0000-0002-8437-0433]{Johannes Staguhn}
\affiliation{NASA Goddard Space Flight Center, 8800 Greenbelt Rd, Greenbelt, MD 20771, USA}
\affiliation{Johns Hopkins University, 3400 N. Charles St, Baltimore, MD 21218, USA}

\begin{abstract}

Recent investigations have demonstrated the potential for utilizing a new observational and data analysis technique for studying the atmospheres of non-transiting exoplanets with combined light that relies on acquiring simultaneous, broad-wavelength spectra and resolving planetary infrared emission from the stellar spectrum through simultaneous fitting of the stellar and planetary spectral signatures. This new data analysis technique, called Planetary Infrared Excess (PIE), holds the potential to open up the opportunity for measuring MIR phase curves of non-transiting rocky planets around the nearest stars with a relatively modest telescope aperture. We present simulations of the performance and science yield for a mission and instrument concept that we call the MIR Exoplanet CLimate Explorer (MIRECLE), a concept for a moderately-sized cryogenic telescope with broad wavelength coverage ($1 - 18\;\mu$m) and a low-resolution ($R\sim50$) spectrograph designed for the simultaneous wavelength coverage and extreme flux measurement precision necessary to detect the emission from cool rocky planets with PIE. We present exploratory simulations of the potential science yield for PIE measurements of the nearby planet Proxima Cen b, showing the potential to measure the composition and structure of an Earth-like atmosphere with a relatively modest observing time. We also present overall science yields for several mission architecture and performance metrics, and discuss the technical performance requirements and potential telescope and instrument technologies that could meet these requirements. 

\end{abstract}


\section{Introduction}
\label{sec:intro}

Cool stars provide an advantage in detecting and characterizing smaller planets, since the fractional planet signal is larger in both primary transit and emission/reflection measurements compared with larger stars.  Additionally, the demographics of planet occurrence reveal that the number of small planets per star increases for cooler stars \citep{Dressing2015}, and since the overall stellar number density increases with decreasing stellar mass (75\% of nearby stars are M dwarfs), cool stars dominate the population of nearby stars with small planets. Even more tantalizing is the fact that rocky planets in the habitable zones (HZ) of cool stars have relatively short orbital periods (4 - 30 days), making them amenable to repeated primary transit and secondary eclipse measurements, as well as phase-resolved measurements of the planetary flux.  

Transit observations place constraints on the atmospheric conditions at the planet's terminator, and are effective in obtaining a first assessment of the bulk characteristics of the atmosphere. Unfortunately, the size of atmospheric spectral features is inversely proportional to the square of the stellar radius and proportional to the planet radius and atmospheric scale height; therefore, due to the cool and dense nature of the atmospheres expected for habitable zone rocky planets, the atmospheric scale height will be quite small. The absorption signature of molecular constituents imprinted on the starlight from a transiting habitable Earth-like planet around an M dwarf would be on the order of 5 - 30 parts per million (ppm) even for the very coolest stars, potentially requiring 10s - 100s of repeated visits with the James Webb Space Telescope (JWST) to detect H$_2$O or other signs of habitability in the atmosphere of TRAPPIST-1e, one of the best transiting temperate terrestrial planets around M-dwarfs for follow-up characterization discovered to date \citep{Morley2017, Wunderlich2019, Fauchez2019, Lustig-Yaeger2019, Suissa2020, Komacek2020}.

Exoplanets can also be characterized using thermal emission. While the signal-to-noise ratio (SNR) for primary transit observations tends to increase towards shorter optical wavelengths where the star is brightest, the SNR for thermal emission increases towards longer infrared wavelengths due to the combined effect of weaker emission from the star and stronger emission from the planet. The SNR peaks in the mid-infrared (MIR) for planets in the HZ. For TRAPPIST-1e, the planet-to-star flux contrast ratio is estimated to be \revision{$<$ 2 ppm} below 7~{\um}, $\sim$25 ppm at 10~{\um}, and $\sim$150 ppm at 20~{\um} \citep{Mullally2019}. However, the SNR for thermal emission measurements is still dependent on the measured brightness of the planet, leading to requirements for numerous secondary eclipse measurements in order to detect the emission from even the most optimal transiting planets in the HZ with JWST \citep{Piette2022}. 

To increase SNR and decrease observing time, we either need a larger collecting area or a brighter target. \citet[henceforth S20]{Stevenson2020} explored the second option, presenting a method for characterizing the thermal emission from non-transiting planets called the Planetary Infrared Excess (PIE) technique.  PIE relies on simultaneous wavelength coverage of spectral regions covering both the planet's infrared flux (longer wavelengths) as well as spectral regions heavily dominated by the star (shorter wavelengths). \citetalias{Stevenson2020} showed that, based on certain simplifying assumptions and sufficient wavelength coverage and measurement precision, the flux from the planet could be extracted from the star by mutually fitting for the thermal flux of both the star and planet. \citetalias{Stevenson2020} performed blackbody simulations of the PIE technique for measurements of the canonical nearby non-transiting rocky exoplanet Proxima Centauri b \citep{Anglada2016} with both JWST and a 2-m MIR telescope, finding that JWST is unlikely to be able to constrain the planet's surface temperature or radius due to insufficient wavelength coverage beyond 13~{\um}.  Conversely, for a 2-m aperture with 3--20~{\um} wavelength coverage and 100 hours of integration time, \citetalias{Stevenson2020} was able to constrain the simulated planet’s temperature to within {$\pm$}15 K and its radius to within {$\pm$}0.1 Earth radii.

\revision{Building on the work from \citetalias{Stevenson2020}, \citet[henceforth LY21]{Lustig-Yaeger2021} utilized realistic planetary atmosphere models to demonstrate that the PIE technique applied to measurements of the hot Jupiter WASP-43b.  Of importance to this work, \citetalias{Lustig-Yaeger2021} found that sampling the Wien edge of the planet spectrum provides the best temperature constraint, while sampling the Rayleigh–Jeans tail provides the best radius constraint, and having both the Wien edge and Rayleigh–Jeans tail drastically reduces the degeneracy between these two planet parameters.}

In this paper we extend the work of \citetalias{Stevenson2020} to examine the instrument and mission requirements for utilizing the PIE technique to characterize the atmospheres of the nearest cool rocky and Neptune-like non-transiting planets around M-dwarf stars, and determine the sample of known planets that could be characterized with different potential aperture sizes.  In Section~\ref{sec:retrieval} we use simulated atmospheric retrievals for PCb to examine the impact of various instrument and mission parameters on the ability to characterize the atmospheric properties of the planet using the PIE technique. In Section~\ref{sec:yields} we scale the results for PCb to examine the total yield of thermal emission measurements of small cool planets that could be achieved in a reasonable mission lifetime \revision{with a stand-alone MIR exoplanet characterization mission concept such as the Mid-IR Exoplanet CLimate Explorer (MIRECLE) concept proposed in \citet{Staguhn2019b}}.  In Section~\ref{sec:trl} we review the technical readiness of the key observatory and instrument sub-systems needed to achieve the performance requirements, and in Section~\ref{sec:discussion} we conclude with a discussion of outstanding issues and future work.

\section{Motivation}
\label{sec:mot}

\begin{figure*}[t]
\centering
\includegraphics[width=\textwidth]{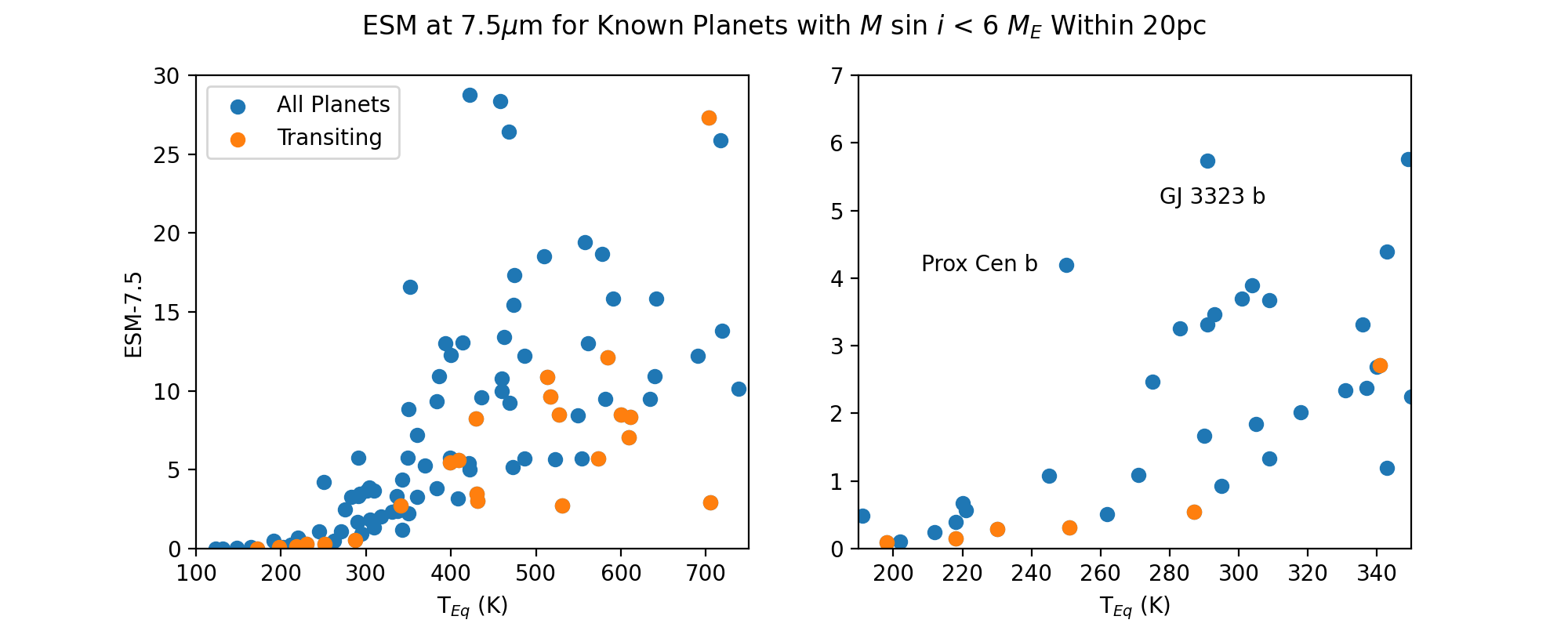}
\includegraphics[width=\textwidth]{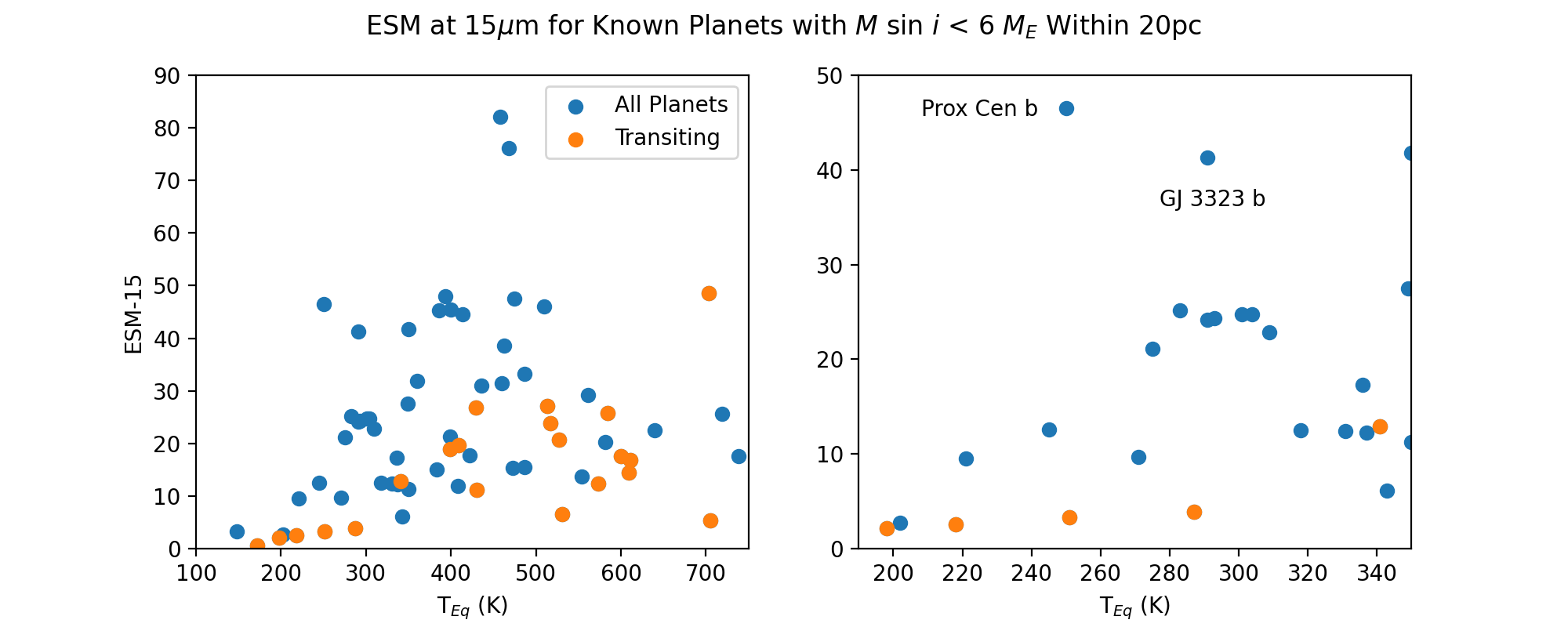}
\caption{Emission spectroscopy metric (ESM) from \citet{Kempton2018} for all known planets with $M~\text{sin}~i$ less than 6 M$_E$. Top: ESM as traditionally defined at 7.5~{\um}.  Bottom: Same metric but defined at 15~{\um}. Transiting planets are marked in orange. Due to the impact of apparent magnitude, the best non-transiting planets have higher ESM, with many low-temperature planets having an ESM at 15~{\um} above the nominal threshold of 10 due to the improved blackbody planet-to-star flux ratio. The two best targets with T$_{eq}<300$K in the Exo.MAST database, Proxima Cen b and GJ 3323 b, are noted.}
\label{fig:ESM}
\end{figure*}

As stated above, we are severely limited by photon noise for thermal emission measurements of transiting HZ rocky planets; however, the opportunities for characterizing HZ rocky planets increase dramatically when non-transiting planets are included as potential targets.  In the top of Figure~\ref{fig:ESM} we plot the emission spectroscopy metric (ESM) from \citet{Kempton2018} for known transiting and non-transiting planets with derived inclination-dependent masses  ($M~\text{sin}~i$) less than 6 M$_{Earth}$ within 20 parsecs.  \revision{Assuming an Earth-like composition (32.5\% Fe and 67.5\% \ce{MgSiO3}), 6 M$_{Earth}$ roughly corresponds to 1.6 R$_{Earth}$, which is a reasonable upper limit in size for terrestrial planets \citep{Rogers2015}.}  The ESM provides a relative metric for comparing the expected SNR for a measurement of a specific planet's thermal emission assuming a fixed observation time, and is determined by the radius and effective temperatures of both the star and planet, the distance to the star, and the reference wavelength of the observation.  The traditional ESM assumes a reference wavelength of 7.5~{\um} and uses a scaling factor that produces a value of 10 for the well-known planet GJ~1132b, which is the best known sub-Neptune-mass transiting planet for emission measurements. It is clear that the addition of non-transiting planets could add a significant number of new targets equivalent to GJ~1132b in their potential for characterization, but even in this scenario, no planet with T$_{eq} < 350$ has an ESM value $> 6$; this is due to the fact that the bulk of their thermal emission is emitted at wavelengths longer than 10~{\um}.  In the bottom of Figure~\ref{fig:ESM} we plot the ESM metric but with a reference wavelength of 15~{\um}.  There are now many targets with ESM $> 10$, with equilibrium temperatures extending down to 220K.

This large number of high-quality targets for the population of non-transiting warm and temperate rocky planets around nearby stars could be investigated at wavelengths beyond 10~{\um}, but would require instrumentation optimized for leveraging the PIE technique at long wavelengths. However, in contrast to secondary eclipse measurements, PIE measurements of time-integrated thermal emission would not be limited in SNR by the duration of the in-eclipse measurement period -- enabling the potential for a relatively small telescope to detect the planetary infrared excess of smaller and cooler planets using long and continuous integration times. 

\citetalias{Stevenson2020} focused on the potential of the PIE technique for constraining the planetary radius and brightness temperature of a non-transiting planet, but PIE spectroscopy in the 8 - 20~{\um} wavelength range would also sample a number of prominent atmospheric absorption features indicative of habitability and/or biology such as \ce{H2O}, \ce{CH4}, \ce{CO2}, \ce{O3}, \ce{NH3}, \ce{N2O}, and \ce{SO2}. \revision{Emission measurements of these atmospheric constituents would provide strong constraints on whether a planet is} a wet, Earth-like planet, a dry, Venus-like planet with a dense \ce{CO2} atmosphere, or a Mars-like planet with a thin \ce{CO2} atmosphere. The strong ozone band at 9.6~{\um} allows for the inference of the presence of molecular oxygen in the atmosphere, which is a powerful biosignature when combined with other out-of-equilibrium molecular species \citep[such as \ce{CH4} and/or \ce{N2O};][]{Schwieterman2018, Thompson2022}. 

Furthermore, long continous integration times enabled by a dedicated mission would also be able to obtain phase-resolved thermal emission measurements (i.e. phase curves). Extending from a single time-integrated thermal emission spectrum to phase-dependent spectroscopy measurements can provide additional constraints on a wealth of atmospheric properties of terrestrial M-dwarf planets.   Phase curves can be used to more firmly distinguish between planets with and without substantial atmospheres \citep[e.g.][]{Maurin2012, Kreidberg2019}, and subsequently identify the presence of hemispheric-wide clouds \citep{Yang2013}. For planets with an atmosphere, phase-resolved thermal emission measurements allow for detailed mapping of a planet's atmospheric composition and thermal structure, which is critical towards assessing  habitability. Phase curves can also constrain the planet’s rotation rate --- the morphology of thermal phase curves can differentiate synchronous rotators with radial or super-rotating winds \citep{Carone2018} from those in spin-orbit resonances \citep{Wang2014}, and can distinguish runaway and non-runaway greenhouse climate states \citep{Kopparapu2017}.

\section{Simulated Atmospheric Retrievals of Proxima Cen b}
\label{sec:retrieval}

The effectiveness of the PIE method is driven by the ability to independently constrain the stellar and planetary contributions to the spectral energy distribution; therefore the key observational factors are the simultaneous wavelength range covered by the instrument and the SNR of the data at different wavelengths.  In order to define the requirements needed to adequately constrain the atmospheric characteristics of temperate rocky planets with a MIRECLE mission, we have conducted atmospheric retrievals on simulated observations of Proxima Cen b, the archetype for the population of non-transiting temperate terrestrial and sub-Neptune planets around the nearest M-stars.  

Proxima Cen b, or PCb, is a planet in the habitable zone of Proxima Centauri, the closest star to our solar system at a distance of 1.3 pc.  It was first detected by \citet{Anglada2016} using RV measurements and subsequently confirmed by \citet{Suarez2020}, with a measured $M\sin(i)$ of 1.27 M$_E$. With an orbital period of 11.2 days and a semimajor axis of 0.049 AU, the planet receives a stellar flux of $0.65-0.7\times$ Earth's insolation (S$_E$) and has a zero-albedo equilibrium temperature corresponding to $\sim270$ K \citep{Turbet2016}. Subsequent to the detection of PCb, several other planet candidates have been tentatively detected in the same system: one similar-mass candidate at a much longer orbital period of 5.21 yr \citep{Damasso2020} and a second much lower-mass candidate with a period of 5.1 days \citep{Faria2022}; however, to date neither candidate has been confirmed. Additionally, a detection of a dust belt around Proxima Cen based on ALMA images was announced by \citet{Anglada2017}, but the signal was later determined to be due to an extreme flare event \citep{MacGregor2018} and no further evidence of dust in the system has been detected.

Based on the close proximity of the target and the potential habitability of the planet, we use PCb as our fiducial target for our proof-of-concept retrieval simulations. To date, essentially nothing is known about the presence and composition of an atmosphere on PCb; however, for the purpose of developing observing techniques and analysis methods capable of assessing the planet's atmospheric characteristics, habitability, and biosignatures, we adopt the common fiducial assumption that the planet is Earth-like in atmospheric composition. Since PCb was discovered by RV and it has not been found to transit \citep{Jenkins2019}, the unknown inclination and planet radius as well as the uncertainty in planet mass mean that the planet could either have a fully rocky composition, a significant ice component, or a H/He envelope (although the latter has been shown to be only 10\% likely; \citet{Bixel2017}). Critically, if PCb is assumed to have formed with a low fraction of H/He, atmospheric loss processes could present a substantial threat to the retention of atmospheres over long timescales and the planet may be airless \citep{Barnes2016, Dong2017, Garcia-Sage2017, Zahnle2017}. On the other hand, the large inherent uncertainties in interior composition propagate into uncertainty in the outgassing rate from the interior \citep[e.g.,][]{Noack2021}, a process that can serve to produce and sustain a secondary atmosphere. Thus the possibility that PCb is a terrestrial planet with an atmosphere cannot be ruled out, and climate and photochemical models of PCb under the assumption of relatively Earth-like atmospheric conditions show that the planet could maintain habitable surface temperatures with stable surface oceans \citep{Turbet2016, Meadows2018, DelGenio2019, Galuzzo2021}. However, it is clear that with the potential additional planet candidates and the impact of continual flaring from the host star, the assumptions underlying these initial simulations may need to be modified in the future - but only observations of PCb can provide definitive answers on the presence and nature of its atmosphere. We further discuss these issues in \S\ref{sec:discussion}. 

In the following subsection \S\ref{sec:retrieval:model} we describe the retrieval model, and then present retrieval results in \S\ref{sec:retrieval:results}. 

\subsection{Retrieval Model} \label{sec:retrieval:model}

We use the Spectral Mapping Atmospheric Radiative Transfer Exoplanet Retrieval (\smarter) model to infer physical parameters from synthetic observed spectra (\citealp{Lustig-Yaeger2020thesis}; \citealp{Lustig-Yaeger2022}; Lustig-Yaeger et al. 2022b, in prep). At the core of the forward model, \smarter uses the Spectral Mapping Atmospheric Radiative Transfer code \citep[\smart;][]{Meadows1996} to perform line-by-line, multi-stream, multi-scattering radiative transfer calculations in a plane-parallel 1D atmosphere. \smart has been used extensively to simulate the emission, reflection, and transmission spectra of solar system planet atmospheres \citep[e.g.,][]{Meadows1996, Robinson2010, Robinson2011, Arney2014, Schwieterman2015b} and exoplanet atmospheres \citep[e.g.,][]{Segura2005, Schwieterman2016, Meadows2018, Lincowski2018}. 

We limit our descriptions of \smart here to the relevant components for modeling the thermal emission of an Earth-like planet from approximately $1-20$~{\um}. As such, we consider atmospheres composed of \ce{CO2}, \ce{O3}, \ce{H2O}, \ce{CH4}, \ce{N2O}, and \ce{N2}. We use the HITRAN2016 \citep{Gordon2017} line list to calculate molecular rotational-vibrational absorption coefficients for input into \smart using the Line-By-Line ABsorption  Coefficient  model \citep[\lblabc;][]{Meadows1996}. We use a nominal Earth composite surface albedo from \citet{Meadows2018}, which results in a surface emissivity of approximately 0.996 throughout the MIR. All radiative transfer calculations with \smart were performed using 8 streams (4 upward and 4 downward; defined at Gaussian points), and Gaussian quadrature integration was used to produce top-of-atmosphere (TOA) thermal emission spectra.    

We adapted a relatively standard exoplanet thermal emission retrieval modeling approach to work with the PIE technique similar to the methods used by \citetalias{Lustig-Yaeger2021} to explore the potential for PIE with JWST. The key difference between standard secondary eclipse thermal emission modeling and PIE modeling is that PIE requires joint modeling of planetary \textit{and stellar} parameters. For PIE, the wavelength-dependent observable flux from the exoplanetary system ($F_{\rm sys}$) is modeled as 
\begin{equation}
\label{eqn:pie_model}
    F_{\rm sys} = F_{\rm s} + F_{\rm p}
\end{equation}
where $F_{\rm s}$ is the flux from the star and $F_{\rm p}$ is the flux from the planet, and both are distance-scaled to the observer at/near Earth. As discussed in \citetalias{Lustig-Yaeger2021}, many additional factors can be added to \Cref{eqn:pie_model} to account for additional flux sources that may be unresolved in the field, such as exozodiacal emissions; however, we neglect these in this work to focus exclusively on the first-order planet-star effects. Our stellar flux model uses Phoenix stellar grids \citep{allard2003model, allard2007k, Allard2012} accessed via the \texttt{pysynphot} code \citep{STScI2013} and linearly interpolated in effective temperature $T_{\rm eff}$, surface gravity $\mathrm{log}(g)$, and metallicity $\mathrm{log}([M/H])$. In addition, we scale the stellar spectrum using the stellar radius $R_{\rm s}$ and the distance $d$, although only $R_{\rm s}$ is used as a free parameter because $d$ is well known for the Proxima Centauri system, so it is held fixed.   

For the planetary flux model, the \smarter forward model assumes evenly mixed volume mixing ratios (VMRs) for the molecules composing the atmosphere. The forward model uses an adiabatic equation of state parameterization for the vertical temperature profile, 
\begin{equation}
\label{eqn:adiabatic_tp}
    T(P)=
    \begin{cases}
    T_0\left(\frac{P}{P_0}\right)^{\alpha(\gamma-1)/\gamma}, ~\mathrm{for}~ P \ge P_1 \\ 
    T_0\left(\frac{P_1}{P_0}\right)^{\alpha(\gamma-1)/\gamma}, ~\mathrm{for}~ P<P_1
    \end{cases}
\end{equation}
where $T(P)$ is the temperature as a function of pressure $P$, $T_0$ is the reference/surface temperature, $P_0$ is the reference/surface pressure, $\gamma$ is the ratio of specific heats, $\alpha$ is used to modify the \revision{dry} adiabat due to, for example, latent heat release or non-constant specific heats \citep{Catling2017, Tolento2019}. \revision{Since $\gamma = c_p/c_v$, it could be calculated in terms of the gases composing the bulk atmospheric volume rather than treated as a free parameter in the retrieval. However, bulk atmospheric compositions for high mean molecular weight atmospheres are not trivially constrained by retrievals \citep{Barstow2020}, so we directly fit for $\gamma$ as a free parameter independent of the retrieved gas VMRs.} 
\revision{Similarly, although reparameterizing \Cref{eqn:adiabatic_tp} to combine $\gamma$ and $\alpha$ into a single parameter may be a sensible choice for reducing the dimensionality of the retrieval posterior sampling problem, we elected to keep both physical parameters and marginalize over their degeneracy to more easily track how the retrieval constrains them individually.} 

For pressures lower than $P_1$ (nominally the tropopause pressure), an isothermal profile is assumed, fixed to the value of the profile at $P_1$. Thus, \Cref{eqn:adiabatic_tp} results in five free retrieval parameters that together describe the vertical TP profile. We note that this TP profile is not capable of producing temperature inversions, such as the stratospheric inversion in Earth's atmosphere, and is therefore limited. However, as we will show, it is capable of fitting the spectrum of, and retrieving unbiased information about, an Earth-like exoplanet's atmosphere using the spectral resolution and precision considered for MIRECLE. 

Given this 1D description of the planetary atmosphere, along with the aforementioned gas optical properties and surface emissivity information, \smart simulates top-of-atmosphere planetary emission spectra, as previously described, which are added to the stellar flux as shown in \Cref{eqn:pie_model}. 

\begin{figure}
    \centering
    \includegraphics[width=0.49\textwidth]{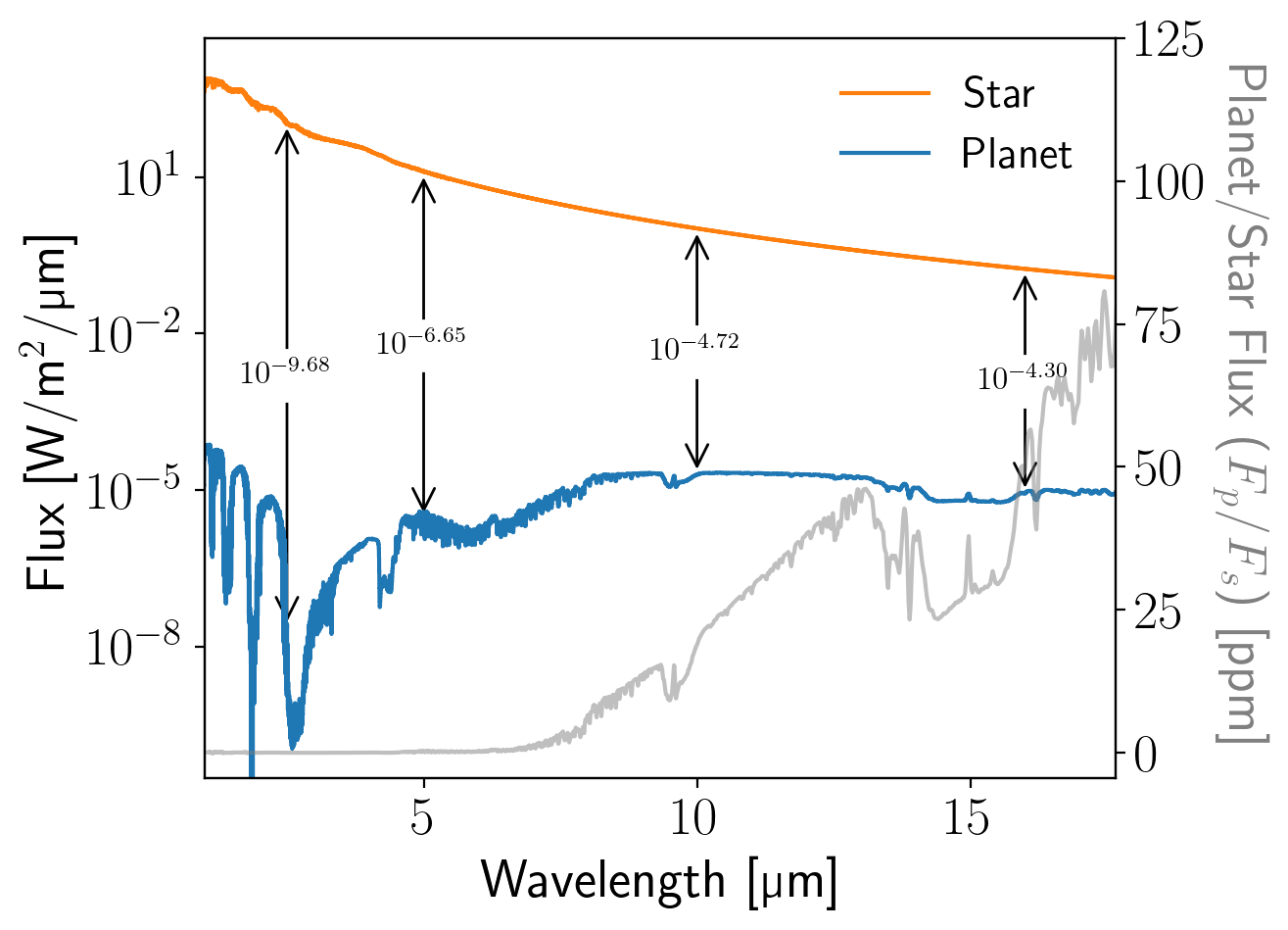}
    \caption{\revision{Simulated stellar and planetary flux for the Proxima Cen system. The left axis shows the relative fluxes of the two sources; the right axis shows the planet-to-star flux ratio. The planetary spectrum peaks at ${\sim}10.4$~{\um}, the stellar spectrum peaks at ${\sim}1$~{\um}, and $F_p/F_s$ steadily rises towards longer wavelengths.}}
    \label{fig:star_planet_flux}
\end{figure}

To produce a plausible observed spectrum of PCb for our retrieval experiments, we used a 1D Earth-like atmospheric model from \citet{Meadows2018} (see their Figure 8) as our fiducial ``truth''. This 1D thermal structure and composition is vertically resolved as a function of altitude, and therefore contains realistic vertical composition gradients, such as a stratospheric \ce{O3} bulge and a \ce{H2O} profile that increases in abundance from the tropopause down to the surface. These \revision{additional vertically-resolved} atmospheric inputs were then provided to \smart to produce a high resolution emission spectrum for PCb \revision{to use as the synthetic data} for our simulated observations. 

\revision{\Cref{fig:star_planet_flux} shows the high-resolution simulated stellar and planetary spectra for reference, prior to degrading them appropriately for use as our simulated observations. The reflected component of the planetary flux declines following the stellar spectrum out to between 3 and 4 \um where the thermal component begins to sharply rise on the Wein tail of the planet’s thermal blackbody. The reflected planet flux contributes ${<}0.1$~ppm to the total system flux at all wavelengths considered and is henceforth neglected. } 

We used the GSFC Planetary Spectrum Generator (PSG; \citet{Villanueva2018}) to produce noise estimates for the PIE observations of PCb under various different assumptions for the MIRECLE telescope and instrument performance. The PSG noise estimation module includes uncertainty contributions from both the astrophysical scene as well as the instrument and telescope.  Specifically, PSG includes photon noise for both the observed object (in this case, star and planet) as well as the astrophysical background flux from the Solar System zodiacal background.  It also includes noise contributions from the thermal background of the telescope and the noise produced by the instrument detectors.

\begin{figure*}[t]
\centering
\includegraphics[width=0.95\textwidth, trim={2.5cm 0 1cm 0}, clip]{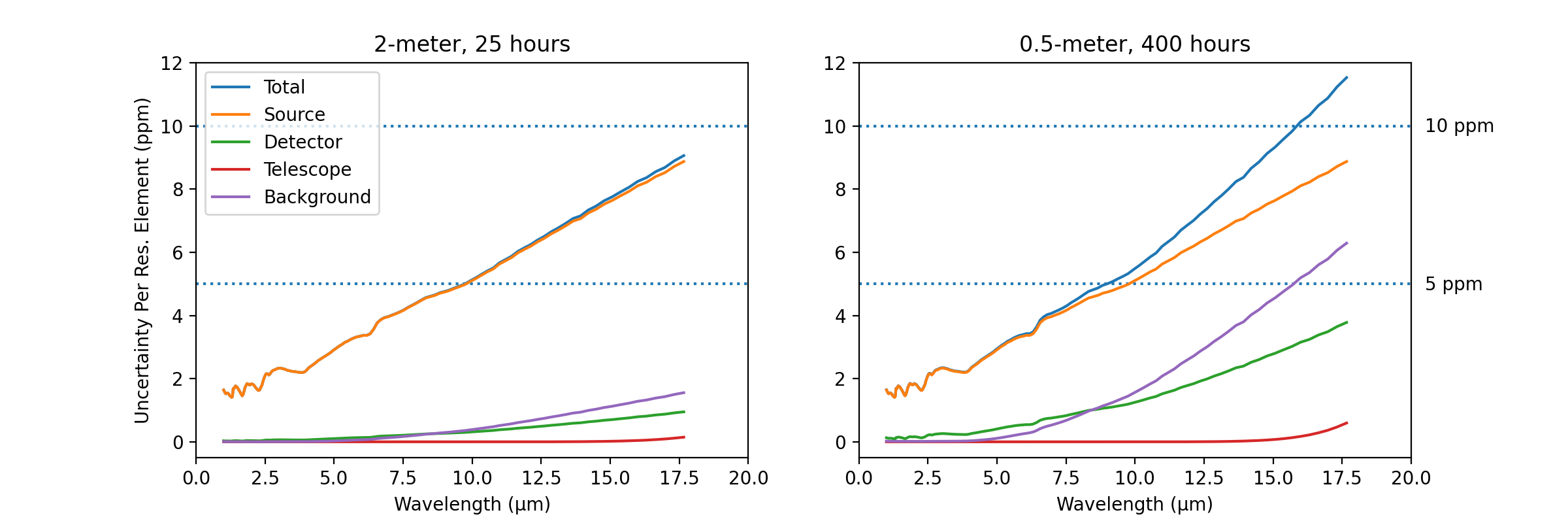}
\caption{Noise contributions for an observation of PCb as a function of wavelength for the 2-m and 0.5-m telescope sizes, assuming the additional parameters listed in Table~\ref{tab:param}. Observations with a 2-m telescope are photon-noise dominated through the wavelength range, while the 0.5-m telescope quickly becomes dominated by the zodiacal background and detector noise. The positions of 5 ppm and 10 ppm noise floors are indicated by dotted lines.} 
\label{fig:noise}
\end{figure*}

PSG utilizes planetary parameters from the NExSci Exoplanet Database. We varied the wavelength range and aperture size based on our assumed mission parameters, and set the instrument aperture size to match $2\times$FWHM at the longest wavelength, in order to avoid significant flux loses.  Since Proxima Cen has a very high ecliptic latitude, we use the minimum value for the Solar System zodiacal light background of 23.3 mag$_V$ per arcsec$^2$, based on the PSG parameterization of the zodiacal flux computed by combining data from \citet{Kwon2004} and \citet{Leinert1998}.  These values and additional parameters for the telescope throughput, primary mirror temperature and emissivity are listed in Table~\ref{tab:noise}. \revision{In Figure~\ref{fig:noise} we plot the noise contributions from the various sources of noise for an observation of PCb, assuming a 2-m aperture versus a 0.5-m aperture. For the 2-m telescope, the photon noise from the star dominates at all wavelengths, and never rises above 9 ppm.  However, for a 0.5-m aperture, the sky background and detector dark current become increasingly important past 10 \um.  For more distant targets or smaller apertures, observations will eventually be background- and detector-limited, leading to longer exposure times to reach the desired SNR.}

\begin{table}
\centering       
\caption{PSG Parameters for Proxima Cen Noise Calculations}
\label{tab:noise}
\begin{tabular}{l | l }     
\hline\hline      
Parameter & Value \\
\hline
Stellar temperature & 3050 K \\
Stellar radius & 0.14 R$_{sun}$ \\
Stellar distance & 1.3 pc \\
Zodiacal background & 23.3 mag$_V$ per arcsec$^2$ \\
Primary mirror diameter & (2 / 1 / 0.5) meters \\
Total throughput & 50\% \\
Telescope temperature & 35 K \\
Detector dark current & 100 e$^{-}$/s \\
Detector read noise & 16 e$^{-}$ \\
Spectral range & (1 / 2.5) - (16.7 / 18)~{\um} \\
Spectral resolution & 50 \\
Spectral aperture size & $2\times$FWHM at LW cut-off \\
Total integration time & 100 hours \\
\end{tabular}
\label{tab:param}
\end{table}

We run our atmospheric retrievals using the \texttt{emcee} Markov Chain Monte Carlo (MCMC) Python package to draw samples from the posterior distribution \citep{Foreman-Mackey2013}. We defined the log-posterior probability function as the sum of a $\chi^2$ log-likelihood function and the log-prior probability. We selected uninformative flat priors for the planetary parameters and Gaussian (and truncated Gaussian) priors for the stellar parameters conditioned using current estimates from the literature for Proxima Cen. \revision{The upper and lower prior bounds on the TP profile parameters $\gamma$ and $\alpha$ correspond, respectively, to a range between 2-10 degrees of freedom for primary molecular constituent of the atmosphere (which is unknown a priori and not guaranteed to be constrained from the spectrum) and an empirical range found within Solar System atmospheres based on the ratio of the true convective lapse rate to the dry adiabatic lapse rate \citep{Catling2017}.} \Cref{tab:retrievals} lists the parameter priors in the third column. We used 30 walkers per dimension, with 14 dimensions unless otherwise stated, yielding a total of 420 walkers in our nominal retrievals. 

We initialized the walkers in a Gaussian ball with mean and standard deviation defined by the results of running a preliminary non-linear least-squares fit to the data using \texttt{scipy.optimize.curve\_fit}. Beginning the MCMC near the maximum \textit{a posteriori} state with the walkers distributed in accordance to an approximation to the 1D marginal posterior distribution allows the MCMC to rapidly burn-in, needing only to identify deviations from Gaussianity and the multi-dimensional parameter covariances. We used \texttt{emcee} to draw samples from the posterior until the length of the chain is equal to approximately $50\times$ the integrated auto-correlation length of the chain---a common convergence heuristic. \revision{While each retrieval runs for a different amount of time, primarily dictated by the wavelength range, for an average forward model evaluation time of 1 minute the full MCMC retrieval running on 25 cores typically takes approximately one month to converge.} 

\subsection{Retrieval Results} \label{sec:retrieval:results}

\begin{figure*}[t]
\centering
\includegraphics[width=0.91\textwidth]{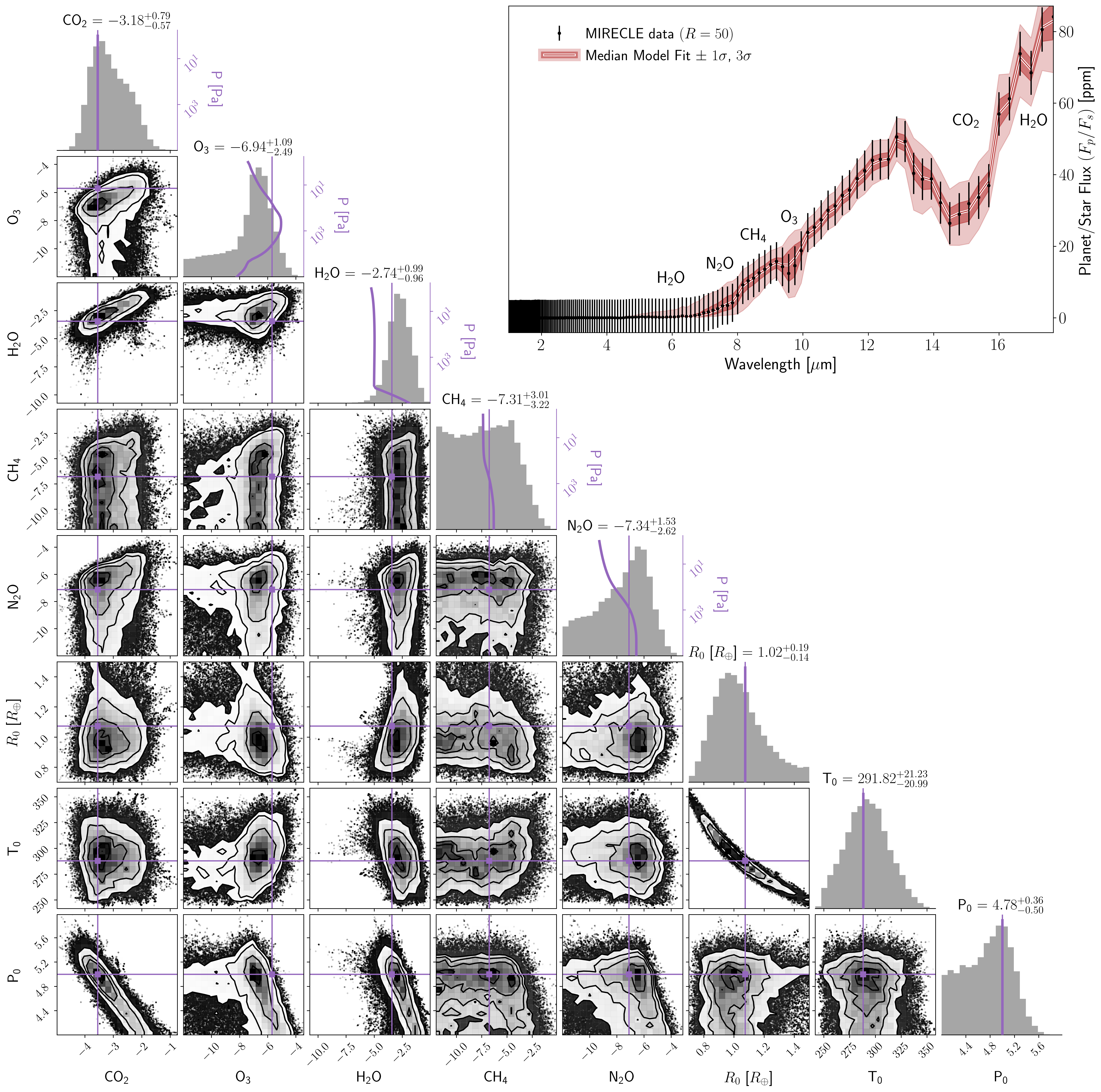}
\caption{Corner plot showing the 1D and 2D marginalized posterior distributions (histograms and contours) for a subset of the retrieved parameters from our fiducial synthetic MIRECLE data. The 1D marginals along the diagonal show the true vertical atmospheric profiles as a function of pressure, from which the synthetic data was generated. The upper right inset shows the median retrieved spectrum sampled from the posterior with envelopes showing the $\pm 1 \sigma$ and $\pm 3 \sigma$ credible intervals. These PIE observations are sufficient to retrieve robust constraints on the VMRs of numerous critical atmospheric gases and the surface temperature.} 
\label{fig:retrieval_corner}
\end{figure*}

For our fiducial retrieval experiment (Case 1) we considered a 2-m diameter mirror with simultaneous wavelength coverage from $1-18$~{\um} and a 5 ppm noise floor. \Cref{fig:retrieval_corner} shows a corner plot with the retrieved posterior distributions for our fiducial noise case. The upper right quadrant of \Cref{fig:retrieval_corner} displays the distribution of planet spectra that were derived from samples from the posterior distribution, which shows the prominent 15~{\um} \ce{CO2} band, the 9.6~{\um} \ce{O3} band, and the location of other bands which are difficult to discern by eye. Only a subset of the retrieved planetary and stellar parameters are shown for clarity and to highlight the most important parameters; this includes the evenly mixed VMRs of \ce{CO2}, \ce{O3}, \ce{H2O}, \ce{CH4}, and \ce{N2O}, the solid body planet radius $R_0$, the surface temperature $T_0$, \revision{and the surface pressure $P_0$}. Additional 

\begin{longrotatetable}
\begin{deluxetable*}{lrl|rrrrrrrrr}
\centerwidetable
\tablewidth{0.98\textwidth}
\tabletypesize{\footnotesize}
\tablecaption{Retrieved constraints on planetary parameters using the PIE technique with MIRECLE \label{tab:retrievals}}
\tablehead{\colhead{} & \colhead{} & \colhead{} & \colhead{\bf Case 1} &  \colhead{\bf Case 2} & \colhead{\bf Case 3} & \colhead{\bf Case 4} & \colhead{\bf Case 5} & \colhead{\bf Case 6} & \colhead{\bf Case 7} & \colhead{\bf Case 8} & \colhead{\bf Case 9}  \\
           \colhead{} & \multicolumn{2}{r}{\bf Wavelengths (\um):} & \colhead{1-18} & \colhead{1-18} & \colhead{1-18} & \colhead{1-18} & \colhead{1-18} & \colhead{1-16.7} & \colhead{2.5-18} & \colhead{5-12} & \colhead{5-12} \\
           \colhead{} & \multicolumn{2}{r}{\bf Diameter (m):} & \colhead{2} & \colhead{2} & \colhead{2} & \colhead{1} & \colhead{0.5} & \colhead{2} & \colhead{2} & \colhead{6.5} & \colhead{6.5} \\
           \colhead{} & \multicolumn{2}{r}{\bf Noise Floor (ppm):} & \colhead{5} & \colhead{0} & \colhead{10} & \colhead{5} & \colhead{5} & \colhead{5} & \colhead{5} & \colhead{5} & \colhead{20} \\
           \colhead{} & \multicolumn{2}{r}{\bf Int. Time (hrs):} & \colhead{100} & \colhead{100} & \colhead{100} & \colhead{100} & \colhead{100} & \colhead{100} & \colhead{100} & \colhead{12.6\tablenotemark{a}} & \colhead{12.6\tablenotemark{a}} \\
           \hline
           \colhead{\bf Parameters} & \colhead{\bf Defaults} & \colhead{\bf Priors} & \multicolumn{9}{c}{\bf Retrieved Constraints (Median$^{+1\sigma}_{-1\sigma}$)} }
\startdata
       \ce{CO2} [VMR]           &   $-3.55$\tablenotemark{b}  & $\mathcal{U}(-12,0)$     &                      $-3.10^{+0.81}_{-0.59}$ &                      $-3.40^{+0.46}_{-0.32}$ &                   $-3.12^{+0.83}_{-0.86}$ &                      $-3.00^{+0.83}_{-0.87}$ &                      $-3.4^{+1.4}_{-1.4}$ &                      $-3.12^{+0.73}_{-0.80}$ &                   $-3.03^{+0.78}_{-0.72}$ &                      $-6.1^{+3.6}_{-4.0}$ &                   $-6.2^{+4.3}_{-3.9}$ \\
        \ce{O3} [VMR]           &   $-5.70$\tablenotemark{b}  & $\mathcal{U}(-12,0)$     &                         $-6.9^{+1.2}_{-2.5}$ &                      $-6.71^{+0.46}_{-0.29}$ &                      $-7.8^{+2.0}_{-2.8}$ &                         $-7.4^{+1.6}_{-3.0}$ &                      $-7.9^{+2.4}_{-2.8}$ &                         $-7.3^{+1.4}_{-2.8}$ &                      $-7.1^{+1.4}_{-3.2}$ &                      $-7.6^{+2.0}_{-3.0}$ &                   $-7.0^{+4.3}_{-3.3}$ \\
       \ce{H2O} [VMR]           &  $-3.44$\tablenotemark{b}   & $\mathcal{U}(-12,0)$     &                         $-2.7^{+1.0}_{-1.0}$ &                      $-2.97^{+0.62}_{-0.48}$ &                      $-4.0^{+2.0}_{-4.5}$ &                         $-3.4^{+1.7}_{-4.1}$ &                      $-4.8^{+2.6}_{-4.1}$ &                         $-3.3^{+1.4}_{-3.1}$ &                      $-2.8^{+1.2}_{-1.5}$ &                      $-5.6^{+3.6}_{-4.0}$ &                   $-5.7^{+3.9}_{-4.1}$ \\
       \ce{CH4} [VMR]           &   $-6.76$\tablenotemark{b}  & $\mathcal{U}(-12,0)$     &                         $-7.5^{+3.3}_{-3.2}$ &                         $-7.1^{+2.0}_{-3.1}$ &                      $-6.7^{+3.7}_{-3.5}$ &                         $-6.9^{+3.2}_{-3.4}$ &                      $-6.8^{+3.7}_{-3.3}$ &                         $-7.3^{+3.0}_{-3.2}$ &                      $-6.7^{+2.8}_{-3.6}$ &                      $-7.3^{+3.2}_{-3.2}$ &                   $-6.3^{+4.2}_{-3.9}$ \\
       \ce{N2O} [VMR]           &   $-7.12$\tablenotemark{b}  & $\mathcal{U}(-12,0)$     &                         $-8.0^{+2.1}_{-2.7}$ &                         $-7.0^{+0.9}_{-2.6}$ &                      $-7.6^{+2.2}_{-2.9}$ &                         $-7.9^{+2.2}_{-2.8}$ &                      $-8.0^{+2.9}_{-2.7}$ &                         $-8.0^{+2.1}_{-2.6}$ &                      $-8.5^{+2.3}_{-2.4}$ &                      $-8.1^{+2.9}_{-2.7}$ &                   $-6.9^{+4.4}_{-3.5}$ \\
        $R_p$ [R$_{\oplus}$]    &   $1.074$                   & $\mathcal{U}(0.7,1.5)$   &                       $1.01^{+0.19}_{-0.14}$ &                    $1.031^{+0.078}_{-0.070}$ &                    $0.91^{+0.29}_{-0.13}$ &                       $0.96^{+0.21}_{-0.14}$ &                    $0.93^{+0.23}_{-0.14}$ &                       $1.06^{+0.20}_{-0.18}$ &                    $0.98^{+0.23}_{-0.14}$ &                    $1.06^{+0.29}_{-0.24}$ &                 $1.06^{+0.29}_{-0.26}$ \\
        $P_0$ [log(Pa)]         &     $5.0$                   & $\mathcal{U}(4,6)$       &                       $4.73^{+0.37}_{-0.50}$ &                       $4.89^{+0.18}_{-0.27}$ &                    $4.62^{+0.53}_{-0.44}$ &                       $4.58^{+0.53}_{-0.41}$ &                    $4.62^{+0.65}_{-0.44}$ &                       $4.67^{+0.45}_{-0.47}$ &                    $4.64^{+0.44}_{-0.46}$ &                    $4.62^{+0.68}_{-0.45}$ &                 $4.93^{+0.67}_{-0.62}$ \\
        $P_1$ [log(Pa)]         &    $4.34$                   & $\mathcal{U}(1.34,5.34)$ &                       $4.08^{+0.44}_{-0.50}$ &                       $4.40^{+0.21}_{-0.31}$ &                    $3.95^{+0.57}_{-0.54}$ &                       $3.94^{+0.56}_{-0.60}$ &                    $3.81^{+0.75}_{-1.15}$ &                       $3.98^{+0.52}_{-0.51}$ &                    $4.02^{+0.48}_{-0.49}$ &                       $3.5^{+1.2}_{-1.4}$ &                    $3.5^{+1.3}_{-1.5}$ \\
        $T_0$ [K]               &   $288.0$                   & $\mathcal{U}(100,400)$   &                            $293^{+22}_{-21}$ &                        $290.6^{+9.5}_{-9.0}$ &                         $304^{+27}_{-33}$ &                            $296^{+24}_{-24}$ &                         $299^{+28}_{-27}$ &                            $285^{+25}_{-20}$ &                         $294^{+23}_{-25}$ &                         $287^{+38}_{-29}$ &                      $269^{+59}_{-77}$ \\
     $\gamma$                   &    $1.45$\tablenotemark{b}                   & $\mathcal{U}(1.2,2.0)$   &                       $1.44^{+0.34}_{-0.19}$ &                       $1.45^{+0.30}_{-0.18}$ &                    $1.44^{+0.33}_{-0.19}$ &                       $1.42^{+0.38}_{-0.18}$ &                    $1.46^{+0.35}_{-0.20}$ &                       $1.41^{+0.36}_{-0.17}$ &                    $1.45^{+0.30}_{-0.20}$ &                    $1.55^{+0.33}_{-0.27}$ &                 $1.56^{+0.31}_{-0.25}$ \\
     $\alpha$                   &     $0.6$\tablenotemark{b}                   & $\mathcal{U}(0.4,0.9)$   &                       $0.57^{+0.20}_{-0.13}$ &                       $0.73^{+0.12}_{-0.09}$ &                    $0.57^{+0.21}_{-0.12}$ &                       $0.58^{+0.20}_{-0.13}$ &                    $0.59^{+0.20}_{-0.14}$ &                       $0.57^{+0.19}_{-0.13}$ &                    $0.57^{+0.20}_{-0.13}$ &                    $0.65^{+0.16}_{-0.17}$ &                 $0.66^{+0.16}_{-0.19}$ \\
\enddata
\tablecomments{$\mathcal{U}$ denotes a uniform prior probability distribution described in terms of the lower and upper bounds.}
\tablenotetext{a}{An integration time of 12.6 hours for a 6.5-meter aperture is equivalent to 100 hours with a 2-m aperture, thereby preserving the photon-noise contribution.}
\tablenotetext{b}{These are approximate (column averaged) values to be used as a reference since the simulated data were generated using vertically resolved profiles from \citep{Meadows2018}.}
\end{deluxetable*}
\end{longrotatetable}

\noindent parameters and their $\pm 1 \sigma$ posterior constraints are listed in \Cref{tab:retrievals} under Case 1. The VMRs of \ce{CO2} and \ce{H2O} are constrained to better than an order of magnitude ($< 1$ dex), \ce{O3} and \ce{N2O} have strongly peaked posteriors with slight tails extending towards low abundances, and the \ce{CH4} constraint is most consistent with an upper limit. There is a curving degeneracy (negative correlation) seen between the surface temperature and planet radius, which is consistent with the PIE retrieval findings from \citetalias{Stevenson2020} and \citetalias{Lustig-Yaeger2021}, and underscores a key challenge that the PIE technique faces for non-transiting planets that do not have well constrained radii. Put simply, an increase in planetary flux across the entire wavelength range could be the result of an increase in the planet radius or an increase in the planet surface temperature; hence the degeneracy. \revision{The surface pressure $P_{0}$ is anticorrelated with the molecular VMRs because an increase in the surface pressure at fixed VMR increases the optical depth. With multiple resolved bands in the spectrum, $P_0$ can be constrained.} 

\begin{figure*}[t]
\centering
\includegraphics[width=0.97\textwidth]{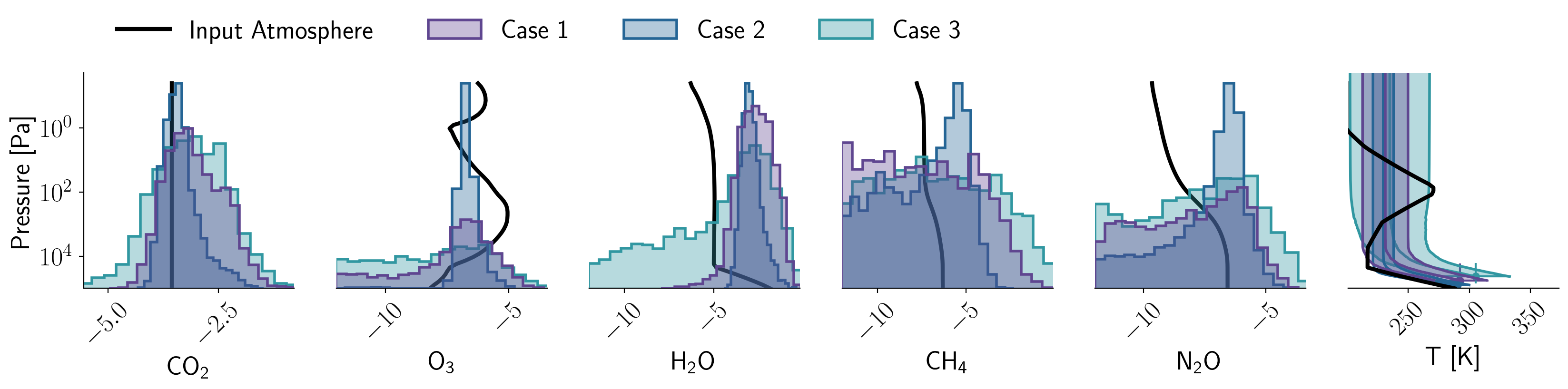}
\includegraphics[width=0.43\textwidth]{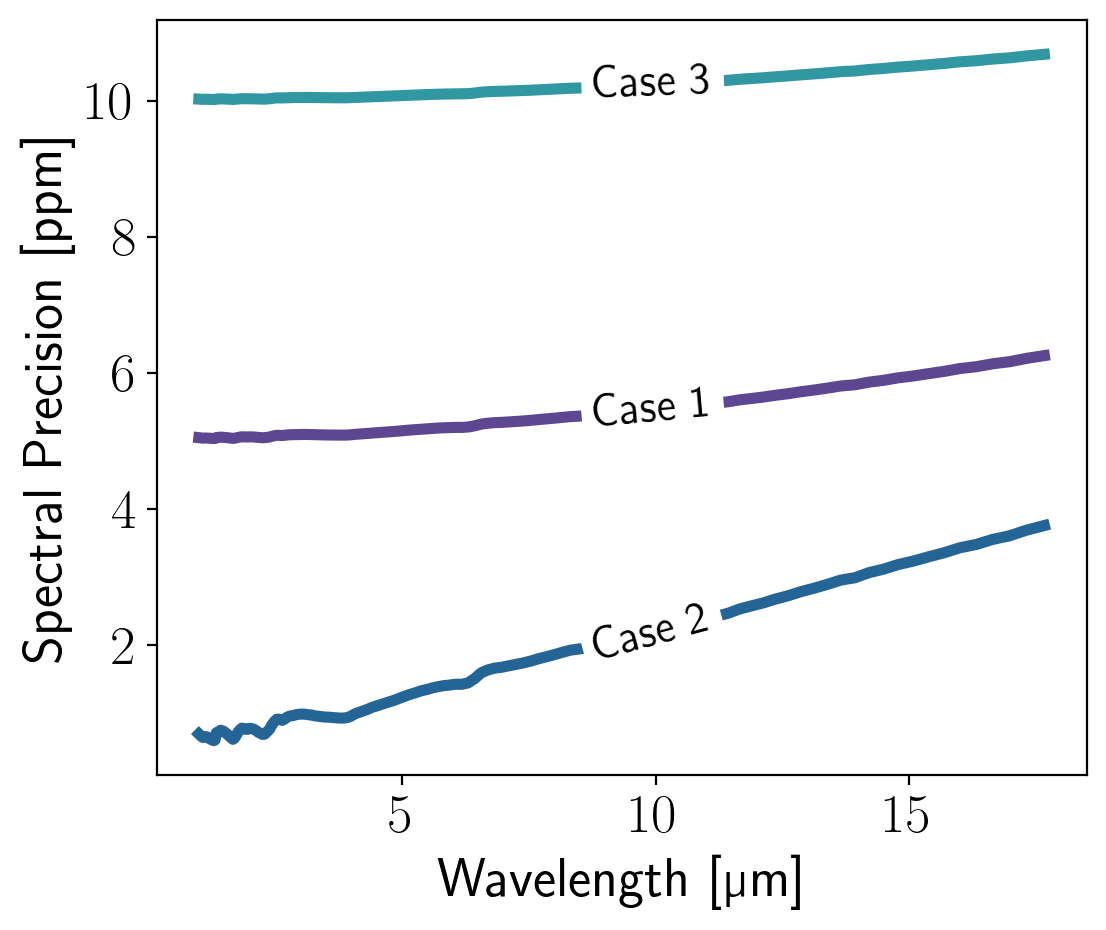}
\includegraphics[width=0.47\textwidth]{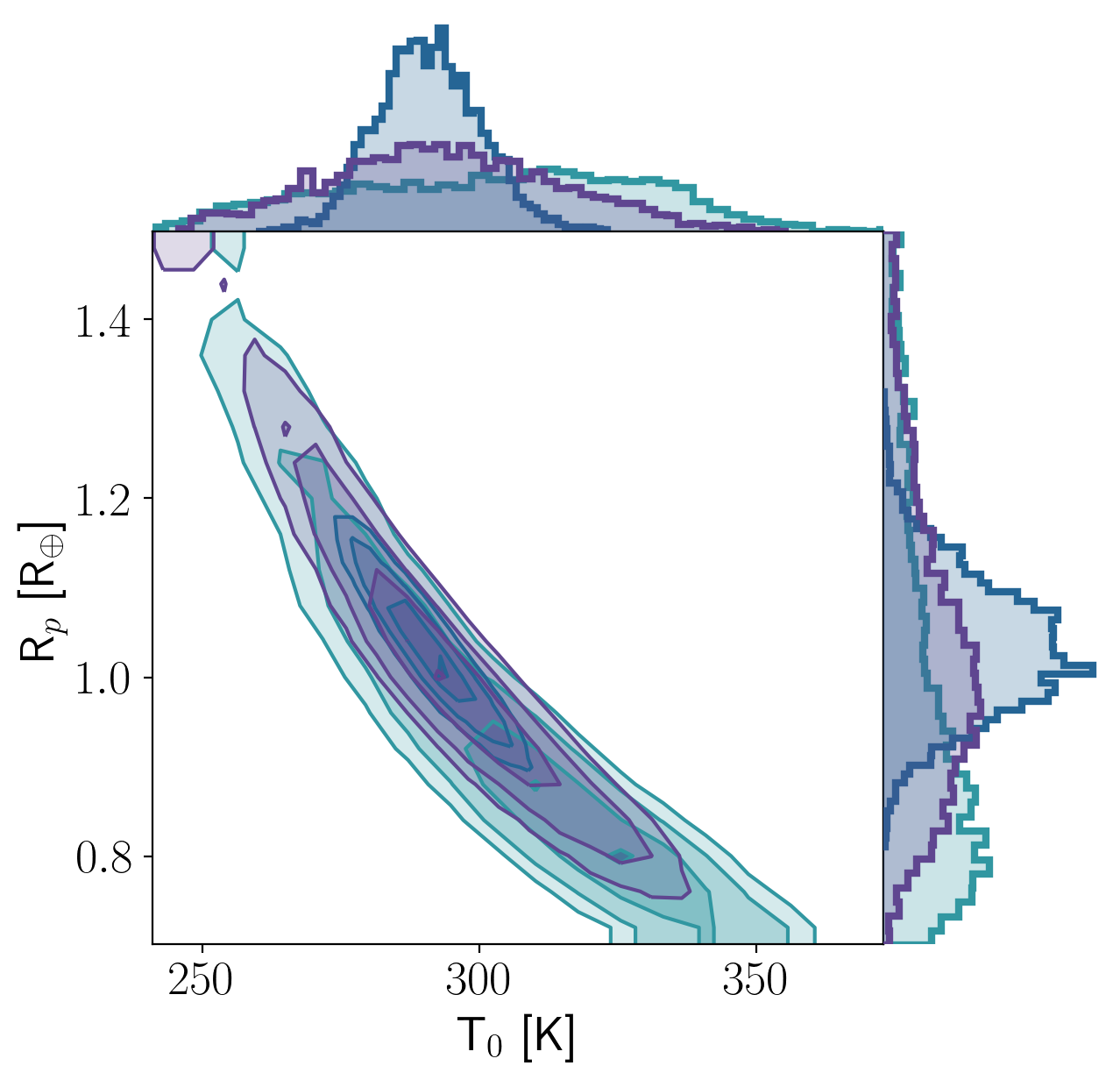}
\caption{Results for retrievals on spectra with varying noise floors. \textit{Top Panels:} 1D marginalized posterior distributions (histograms) for the five gas abundances under consideration (in $\log_{10}$ VMR) and the median retrieved TP profiles with $1 \sigma$ envelopes relative to the true atmospheric profiles (black lines) that were assumed for the input data. \textit{Lower Left:} Spectroscopic precision ($1 \sigma$ errors) as a function of wavelength for the three noise floor cases considered. \textit{Lower Right:} 2D marginalized posterior distributions showing the covariance between surface temperature and planet radius. } 
\label{fig:retrieval_floors}
\end{figure*}

\paragraph{Noise Floor} In our second retrieval experiment we studied how the presence and value of a spectroscopic noise floor affects the atmospheric constraints that can be inferred for PCb using the PIE technique. To complement the Case 1 retrieval that used a 5 ppm noise floor, we repeated the aforementioned experiment using data with no noise floor (Case 2) and data with a 10 ppm noise floor (Case 3). The $\pm 1 \sigma$ posterior constraints for Case 2 and Case 3 are listed in \Cref{tab:retrievals} for reference. The spectroscopic precision of these cases are shown as a function of wavelength in the lower left panel of \Cref{fig:retrieval_floors}. The 1D marginal posterior distributions for the primary atmospheric parameters are compared in the upper row of subpanels in \Cref{fig:retrieval_floors}, and the 2D covariance between the surface temperature and planet radius is shown in the lower right panel of \Cref{fig:retrieval_floors}. As expected, the use of a detector without a noise floor enables observations at the photon limit, which provides the best constraints on the atmosphere. The 5 ppm noise floor case still enables robust constraints on the VMRs of \ce{CO2}, \ce{H2O}, and \ce{O3}. The 10 ppm noise floor case shows significantly degraded sensitivity to the \ce{O3} and \ce{H2O} VMRs, but it is still able to place constraints on the \ce{CO2} abundance and the surface temperature. This comparison also helps to illuminate the nature of the radius-temperature degeneracy, since the photon limited observations are still subject to the same curving negative correlation, but generally show accurate 1D marginals. This indicates that precise observations help to provide unique constraints on each parameter, despite the degeneracy. This is consistent with the fact that temperature and radius are not perfectly degenerate: temperature induces a wavelength-dependent observable (as a blackbody), while the radius is a purely gray effect (modulo the photospheric radius correction discussed in \citet{Fortney2019}, which is not considered in this work).    

\begin{figure*}[t]
\centering
\includegraphics[width=0.97\textwidth]{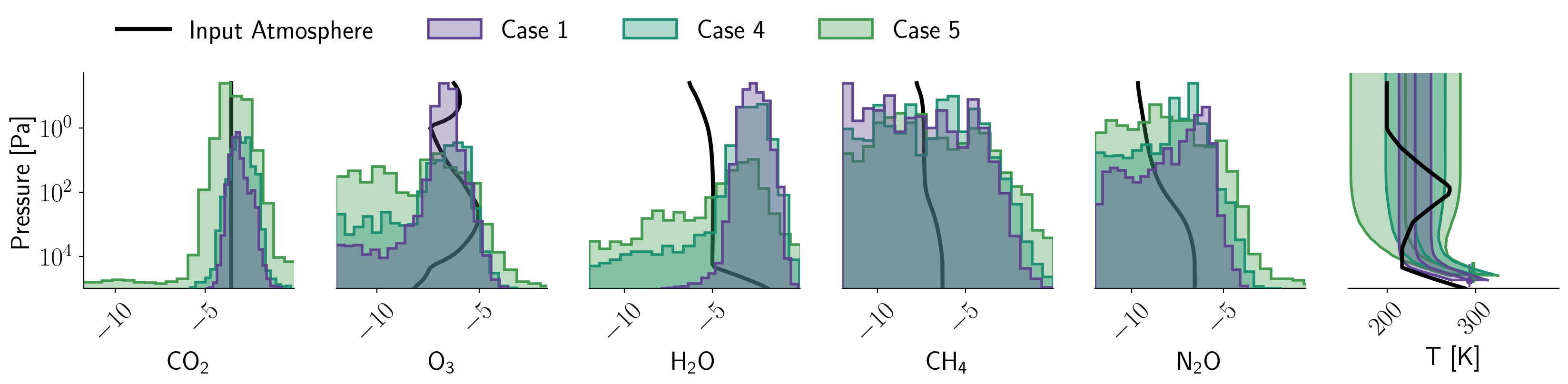}
\caption{Posterior constraints resulting from retrievals on spectra obtained with different primary mirror diameters for the MIRECLE concept. Results using 2-m (Case 1), 1-m (Case 4), and 0.5-m (Case 5) primary aperture sizes are shown in purple, dark green, and light green, respectively. The rightmost panel shows the median retrieved TP profiles and $1 \sigma$ envelopes for each case. } 
\label{fig:retrieval_apertures}
\end{figure*}

\paragraph{Aperture Size} In our next retrieval experiment we kept the noise floor fixed at 5 ppm and examined the effect of aperture size. \revision{A larger aperture size results in both a higher SNR for the same integration time as well as a lower astrophysical background level, since the PSF at a specific wavelength would be smaller and therefore the sky aperture could be smaller as well; this effect can be seen in Figure~\ref{fig:noise}.} We generated synthetic observations assuming a primary mirror diameter of 1 meter (Case 4) and 0.5 meters (Case 5) to compare with our previous results using a 2-m primary mirror diameter (Case 1). For each aperture size we kept the exposure time fixed at 100 hours. The posterior constraints for Case 4 and Case 5 are provided in \Cref{tab:retrievals}. \Cref{fig:retrieval_apertures} compares a subset of the retrieval results. Although the surface temperature ($T_0$) and pressure ($P_0$) constraints are not significantly improved for the 2m compared to the 0.5-m and 1-m (approximately $1.3 \times$ and $1.1 \times$ more constrained at $1\sigma$, respectively), the higher quality data enable $2.0\times$ and $1.2\times$ tighter constraints on the tropopause pressure ($P_1$) at $1 \sigma$, respectively. This is also revealed by the tighter constraints on the retrieved TP profile shown in the rightmost panel of \Cref{fig:retrieval_apertures}. In a 100-hour observations, the 0.5-m telescope would be limited to only detecting the surface/brightness temperature and the presence of \ce{CO2}. The 2-m telescope may be required to provide a robust \ce{H2O} detection, which is critical for assessing habitability. The smaller aperture 1-m and 0.5-m telescopes only approach the 5 ppm noise floor at the very shortest wavelengths (in 100 hours) and as a result, provide comparable constraints on the stellar parameters using the PIE technique, despite these telescope's divergent capabilities at constraining PCb's atmospheric conditions using the data from longer wavelengths. 

\begin{figure*}[t]
\centering
\includegraphics[width=0.97\textwidth]{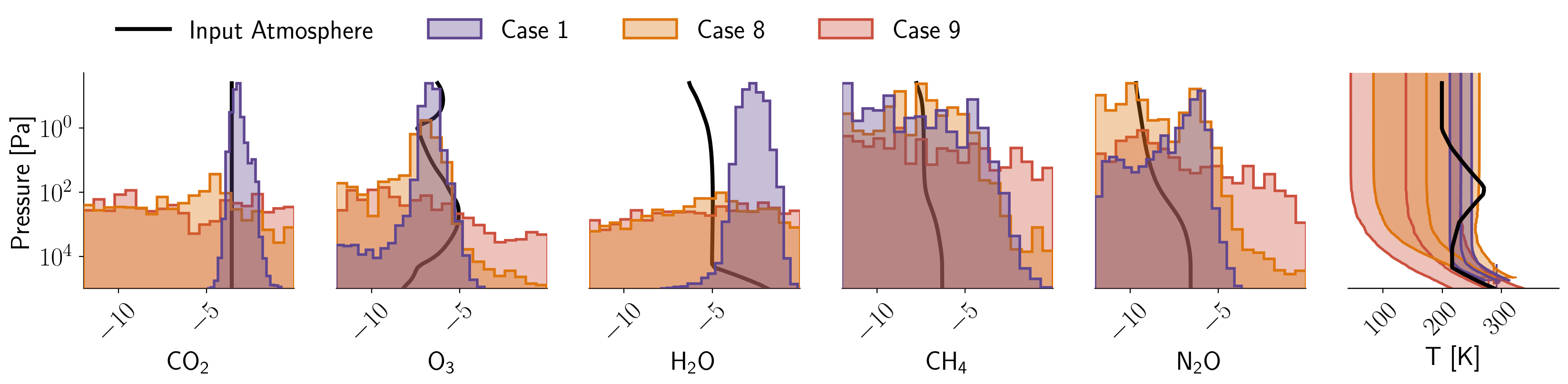}
\caption{Posterior constraints resulting from retrievals on spectra obtained with a 2-m MIRECLE concept (purple), compared to results using the JWST MIRI LRS wavelength range with a MIRECLE-like 5 ppm noise floor (orange), and using JWST MIRI LRS but assuming a 20 ppm noise floor based on MIRI performance expectations. The rightmost panel shows the median retrieved TP profiles and $1 \sigma$ envelopes for each case. Although MIRECLE requires longer exposure times, the combination of broader wavelength coverage and high instrument sensitivity is necessary to enable atmospheric characterization of PCb with PIE.} 
\label{fig:retrieval_wavelength}
\end{figure*}

\paragraph{Wavelength Range}  
In our final retrieval experiment we held the spectroscopic precision (approximately) fixed near a 5 ppm noise floor and varied the wavelength range of the contiguous bandpass to compare different detector capabilities and to perform a comparison with JWST. We investigated five cases: (1) our nominal MIRECLE concept with full wavelength coverage between $1-18$~{\um} (Case 1), (2) a MIRECLE concept with a longwave cutoff at 16.7~{\um} (Case 6; $1-16.7$~{\um}), a MIRECLE concept with a shortwave cutoff at 2.5~{\um} (Case 7; $2.5-18$~{\um}), (4) observations using the same wavelength coverage as JWST's MIRI LRS \citep[$5 - 12$~{\um},][]{Kendrew2015} but using a MIRECLE-like noise floor of 5 ppm (Case 8), and finally JWST's MIRI LRS but assuming a 20 ppm noise floor based on optimistic MIRI performance expectations (Case 9). For our MIRECLE concepts we used an exposure time of 100 hrs. For MIRI LRS we used an exposure time of 12.6 hrs, which results in a similar number of photon counts to the MIRECLE concepts. The $\pm 1 \sigma$ posterior constraints for all of the cases are provided in \Cref{tab:retrievals}.

The loss of a small portion of the wavelength range at the red end of the spectrum (Case 6) and at the blue end (Case 7) yields planetary constraints that are only marginally worse compared to the full $1-18$~{\um} range (Case 1), with one notable exception. The \ce{H2O} constraint swells by a factor of $2.3\times$ for Case 6, indicating that the water absorption seen in \Cref{fig:retrieval_corner} at wavelengths longer than the 15~{\um} \ce{CO2} band (between about $16-18$~{\um}) contributes significantly to the \ce{H2O} abundance retrieval---particularly for ruling out low \ce{H2O} abundances. 

\Cref{fig:retrieval_wavelength} compares the retrieval results for the MIRI LRS cases (Case 8 and Case 9) with our fiducial MIRECLE concept (Case 1). Despite the prominence of \ce{CO2} bands in the IR, the MIRI LRS wavelength range misses the 4.3~{\um} band and all but the shortwave wings of the 15~{\um} band. As a result, MIRI LRS is only weakly sensitive to atmospheric \ce{CO2} in our simulations, as demonstrated in Cases 8 and 9. While MIRI LRS with an unrealistic 5 ppm noise floor shows some sensitivity to the 9.6~{\um} \ce{O3} band, the increase in background noise with wavelength in the MIRI LRS bandpass makes the abundance constraint quite limited. With a 20 ppm noise floor the MIRI LRS sensitivity to \ce{O3} is lost altogether. On the shortwave end, the spectral coverage between $1-5$~{\um} enables MIRECLE to retrieve ${\sim}20\times$ more precise stellar parameters than using the PIE technique with MIRI LRS. 

\revision{In considering the three MIRECLE cases (1, 6, and 7) with 5 ppm noise floors, we see a marginal increase in the retrieved uncertainties on the planet radius and temperature with decreasing wavelength range.  The uncertainties grow noticeably for Case 8, however, where the wavelength range drops to 5 -- 12~{\um}.  The larger uncertainties for Case 8 are to be expected based on results from \citetalias{Lustig-Yaeger2021}.}

We also note that although it may seem that a better comparison would be to have MIRI LRS observe for 100 hours, the longer exposure time would yield negligible gains because the noise floor is already reached in the 12.6 hour integration, so additional integration time would only push up against that floor. Furthermore, MIRI LRS is expected to fully saturate on observations of Proxima Centauri between approximately $5-6$~{\um} \citep{Glasse2015, Kreidberg2016}, which would further decrease the precision of stellar and planetary retrievals using the PIE technique compared to our simulations that neglected saturation. 

\begin{figure*}[t]
\centering
\includegraphics[width=0.97\textwidth]{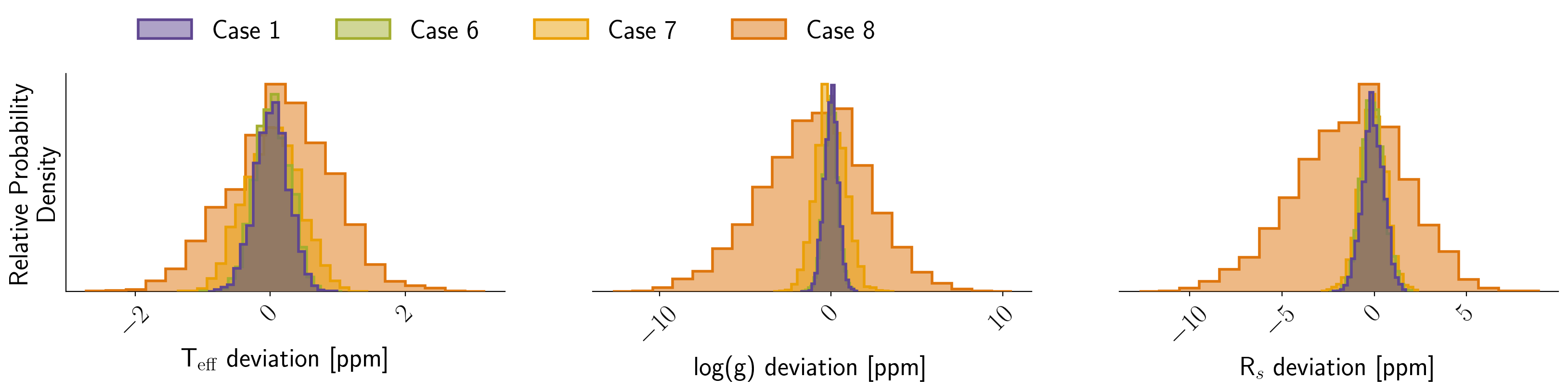}
\caption{Posterior constraints on stellar parameters expressed as deviation from the input value in parts-per-million (ppm) from PIE retrievals using different simulated datasets. The nominal 2-m MIRECLE concept covering $1-18$~{\um} (Case 1) is in purple, a MIRECLE concept covering $1-16.7$~{\um} (Case 6) is in light green, a MIRECLE concept covering $2.5-18$~{\um} (Case 7) is in light orange, and a MIRECLE concept with the JWST MIRI LRS wavelength range of $5 - 12$~{\um} (Case 8) is in dark orange. } 
\label{fig:retrieval_stellar}
\end{figure*}

\Cref{fig:retrieval_stellar} shows the retrieved constraints on the stellar effective temperature $T_{\rm eff}$, stellar surface gravity $\log(g)$, and stellar radius $R_s$ for the cases discussed in this section with different wavelength ranges. Each parameter is shown in ppm deviation from the assumed input value. In general, the constraints on the stellar parameters scale as one would expect. The broadest wavelength range, highest precision observations (Case 1) provide the tightest constraints. Spectra that maintain the same shortwave cutoff, but do not extend as far into the MIR (Case 6) provide effectively the same constraint due to the NIR peak of the stellar blackbody curve. Spectra that do not cover as much of the NIR (Case 7 and Case 8) do not provide as tight of constraints on the stellar parameters. 

However, the stellar parameters are exceedingly well constrained in all cases and reveal a caveat of our simple stellar model parameterization. Since we assumed that the star was a static source over the entire duration of our simulated long-exposure observations, and we sampled our observations from our simple stellar model, we were able to infer exactly what we put in. While this helps to demonstrate that the combined stellar and planetary spectra are separable, realistic observations will have numerous sources of variability that operate on different timescales and dominate at different wavelengths. Moreover, existing stellar models need not be good fits to such high SNR observations. More work is needed to study the effect of more realistic stellar sources and the cost-benefit analysis of including more sophisticated stellar spectral models in PIE retrievals. We note that the PIE technique only requires the stellar parameters to be retrieved in so far as they control the stellar spectrum and allow it to be fit precisely enough to reveal the planet spectrum in the residuals. If increasing the sophistication of the stellar model with more parameters causes the precision on the stellar parameters to degrade via degeneracies between various stellar parameters, the PIE technique can tolerate this. If additional stellar parameters prove degenerate with the planet parameters, then it may become more difficult to separate the planetary and stellar spectra, and uncertainty may propagate into the planetary inference. \revision{Similarly, if stellar models fail to fit the stellar component of the combined light spectra at wavelengths where the planet spectrum is recoverable with PIE (e.g., due to missing stellar opacities or contributions from photospheric heterogeneity or variability), then this is likely to bias the retrieved planet parameters. Thus, the accuracy of stellar spectral models at NIR and MIR wavelengths is critical to the success of PIE, particularly for small terrestrial exoplanets. We discuss avenues for future progress on these issues in \S\ref{sec:sar}.}  

\section{Target Sample Demographics}
\label{sec:yields}

While PCb is one of the most promising potentially rocky planets in the habitable zone of a nearby star, Figure \ref{fig:ESM}  demonstrates that there are a large number of high-quality targets discovered around nearby stars by RV surveys that a PIE survey mission would be able to characterize.  In order to determine the planetary demographics of a potential MIRECLE mission yield, we utilized the Exo.MAST catalog \citep{Mullally2019} to construct an initial target list based on the results from \S\ref{sec:retrieval} and realistic assumptions about the mission lifetime.

We first calculated the exposure time for every planet known around a star within 20 pc, based on scaling the nominal observing time needed to characterize PCb for different aperture sizes (100 hours for 2- and 1-m apertures, 400 hours for a 0.5-m aperture in order to push down to the 5 ppm noise floor).  We used 15~{\um} as a reference wavelength for these calculations, and scaled the planetary signal based on the planetary equilibrium temperature and the stellar photon noise based on the star's effective temperature and K-band magnitude. We used the values in Exo.MAST from the NExSci Exoplanet Archive; we note that the values for planet radius for non-transiting planets in the archive are calculated using the mass-radius relationship from \citet{Chen2017}. 

We then adjusted the SNR values to include the noise contribution from the Solar System zodiacal background. Using iterative noise calculations from PSG, we determined that a 2-m telescope is background-limited for K $> 8.8$, while the background-limited regime is K $> 7.25$ for a 1-m and K $> 5.7$ for a 0.5-m. For potential targets dimmer than the background-limited magnitude, we scaled the SNR based on a background-limited noise assumption rather than a photon-limited assumption.

We also identified all of the multi-planet systems where multiple planets could be expected to produce similar thermal fluxes at 15~{\um} and removed them from the target sample. Since the PIE technique does not remove the signal from additional astrophysical sources in the scene, several planets with similar flux levels could lead to confusion and difficulty in retrieving the atmospheric characteristics of a single planet.  Factors such as the orbital period could be used to disentangle the signals, but for this study we removed planetary systems with multiple planets \revision{where the difference in expected thermal flux between the planets at 15~{\um} is expected to be less than a factor of 10}. The one notable exception to this removal is Proxima Cen, where a third planet candidate interior to PCb was recently announced (see \S\ref{sec:retrieval}); however, since all of these calculations were performed prior to the announcement, and the candidate has not yet been confirmed, we kept PCb in our target list.
 
We then performed cuts in the potential target list by defining a mission goal of examining the transition in planetary properties from rocky planets ($M_{p} < 6 M_{E}$) to Neptune-like planets ($M_{p} < 20 M_{E}$), and planets with potential surface water conditions ($T_{p} < 400K$) versus warmer planets ($400K < T_{p} < 700K$). This regime is also similar to the inner Solar System conditions, and also where JWST will be more limited due to the 13~{\um} wavelength cut-off and the limitations for longer-duration phase curves.  

Finally, we defined two separate assumptions for a primary survey strategy:
\begin{itemize}
    \item A time-integrated thermal emission reconnaissance survey, where the time-average thermal emission spectrum for a planet would be measured based on expectations for SNR calculated from the stellar and planetary parameters
    \item A four-quadrant hemisphere mapping survey, where the minimum observing time is equivalent to the orbital period, and the target SNR needs to be achieved in 1/4 of the total time on-source so variations in phase-dependent emission could be detected
\end{itemize}

In both cases we limit the total observing time to a maximum of 18 months by varying the maximum exposure time, in order to achieve a prime mission period of 2 years with a 75\% duty cycle.  We address the results of both of these reference mission simulations below.  

\subsection{Yields From A Time-Integrated Thermal Emission Survey}
In this scenario, we calculate the total observing time for each target that would be needed to detect an emission spectrum by binning up all the observations, assuming an atmospheric temperature equivalent to the planetary equilibrium temperature.

Our results are summarized in Figure~\ref{fig:emission}, and the full lists of potential targets for each mission scenario are listed in the Appendix. We find that with 18 months of exposure time, a 2-m telescope could obtain spectra for 37 planets, a 1-m telescope could observe 36 planets, and a 0.5-m telescope could examine 20 planets.  The planet temperatures range from 250K to 690K, and inclination-dependent planet masses range from 1.27 M$_E$ to 20 M$_E$ for all mission sizes; the main difference is the total number of planets characterized and the exposure time required for each planet.  The 2-m and 1-m telescope sizes could gather spectra for almost every planet in less than 600 hours, while the 0.5-m telescope would require longer exposure times.

We also conclude that the 2-meter telescope is essentially target-limited in terms of time-integrated thermal emission measurements; even with an exposure time that is 1/4 of the original exposure time (essentially achieving the same sensitivity as a 1-m telescope), the total number of observable planets only rises by 2; all other known targets require significantly long exposure times due to stellar radii or distance. However, if we accept this penalty in SNR, the total required observing time for integrated emission measurements of 39 planets would be only 8 months instead of 18.

\begin{figure*}[t]
\centering
\includegraphics[width=0.95\textwidth]{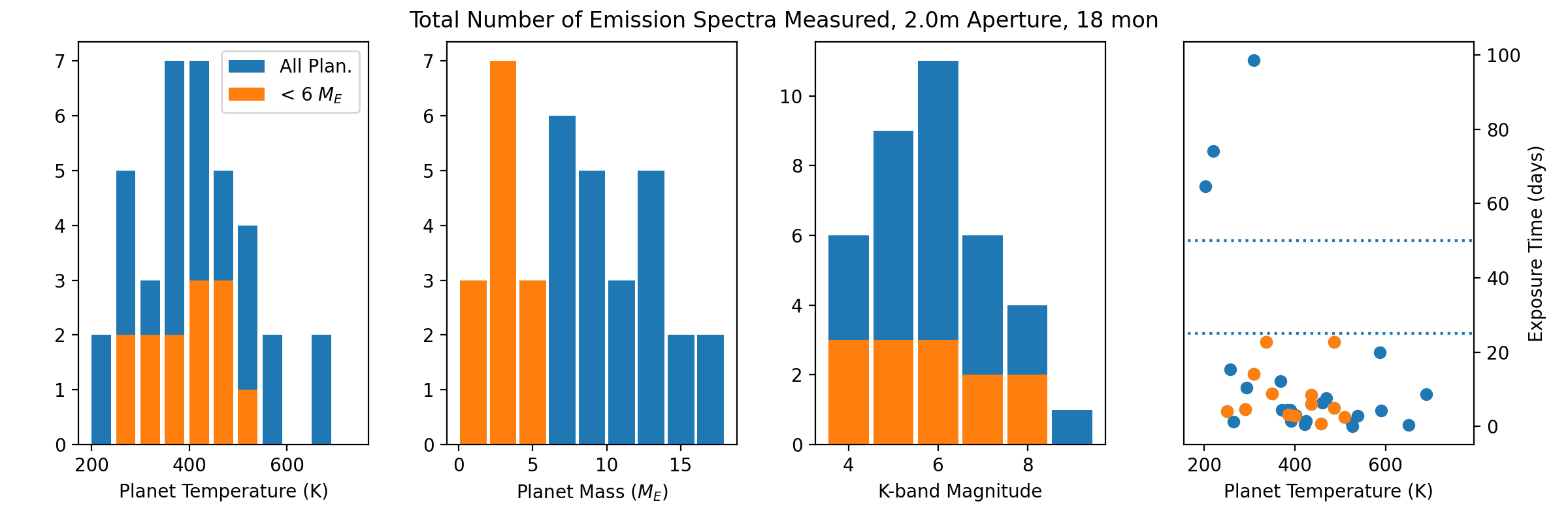}
\includegraphics[width=0.95\textwidth]{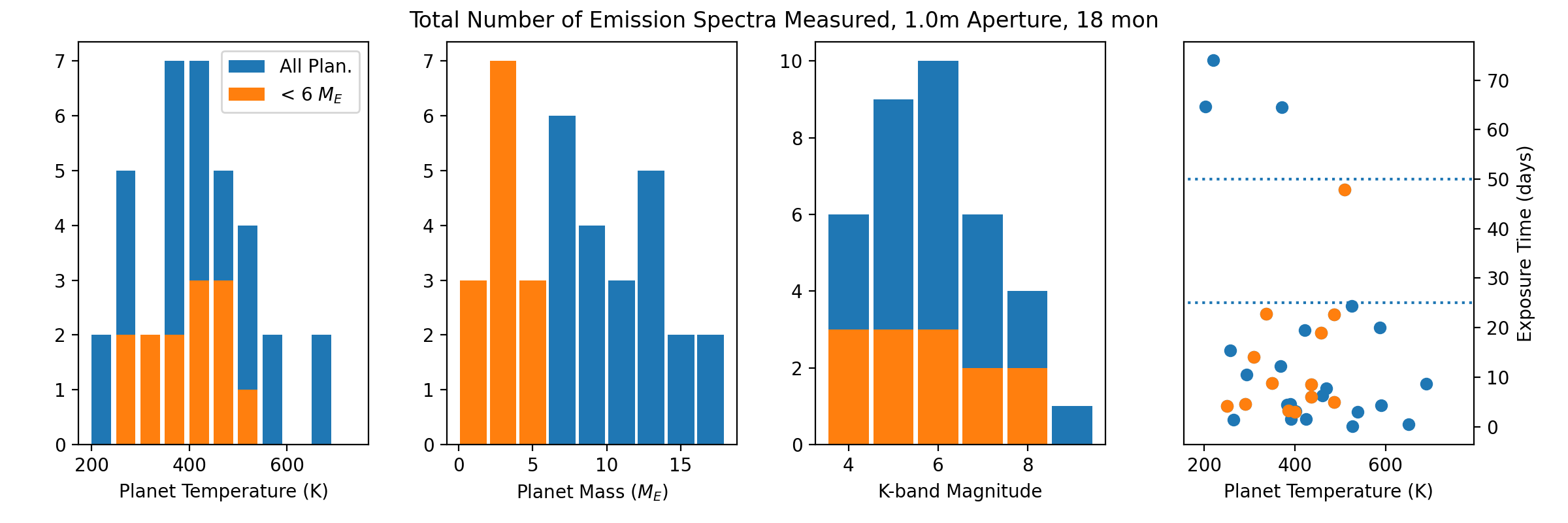}
\includegraphics[width=0.95\textwidth]{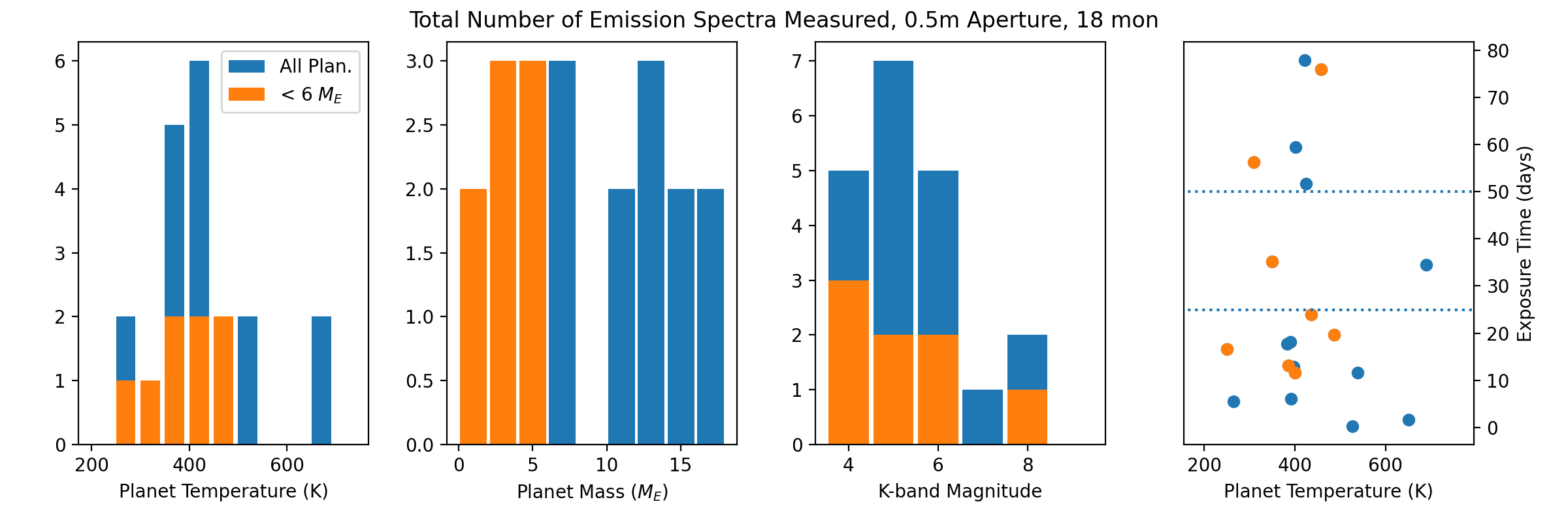}
\caption{Distribution of planet equilibrium temperatures, minimum masses, K-band magnitudes of the host stars, and total exposure times for a survey of time-integrated emission measurements.  The orange bars denote potentially rocky planets ($M_{p} < 6 M_{E}$).  Top: 2-meter aperture, Middle: 1-m, Bottom: 0.5-m.}
\label{fig:emission}
\end{figure*} 


\subsection{Yields From A Thermal Phase Curve Survey}
In this scenario, we calculate the total observing time for each target that would be needed to achieve a four-quadrant phase-resolved measurement - essentially obtaining integrated spectra for the day-side, night-side and two transitional hemispheres, assuming an atmospheric temperature equivalent to the planetary equilibrium temperature.  

\begin{figure*}[t]
\centering
\includegraphics[width=0.95\textwidth]{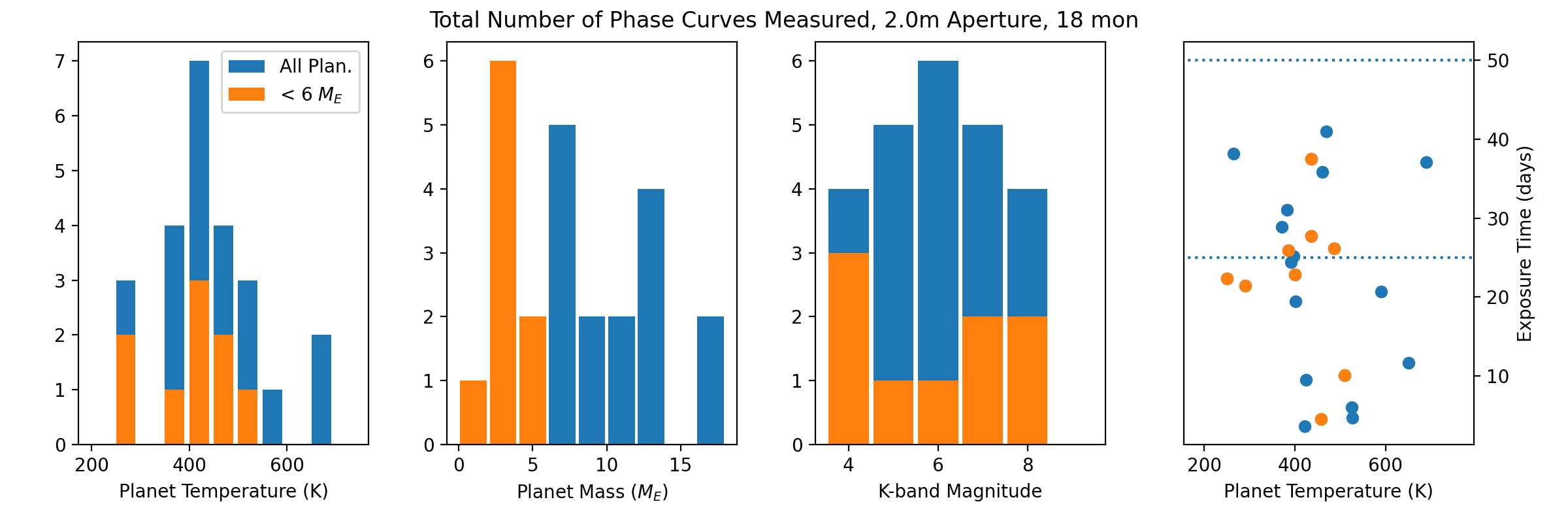}
\includegraphics[width=0.95\textwidth]{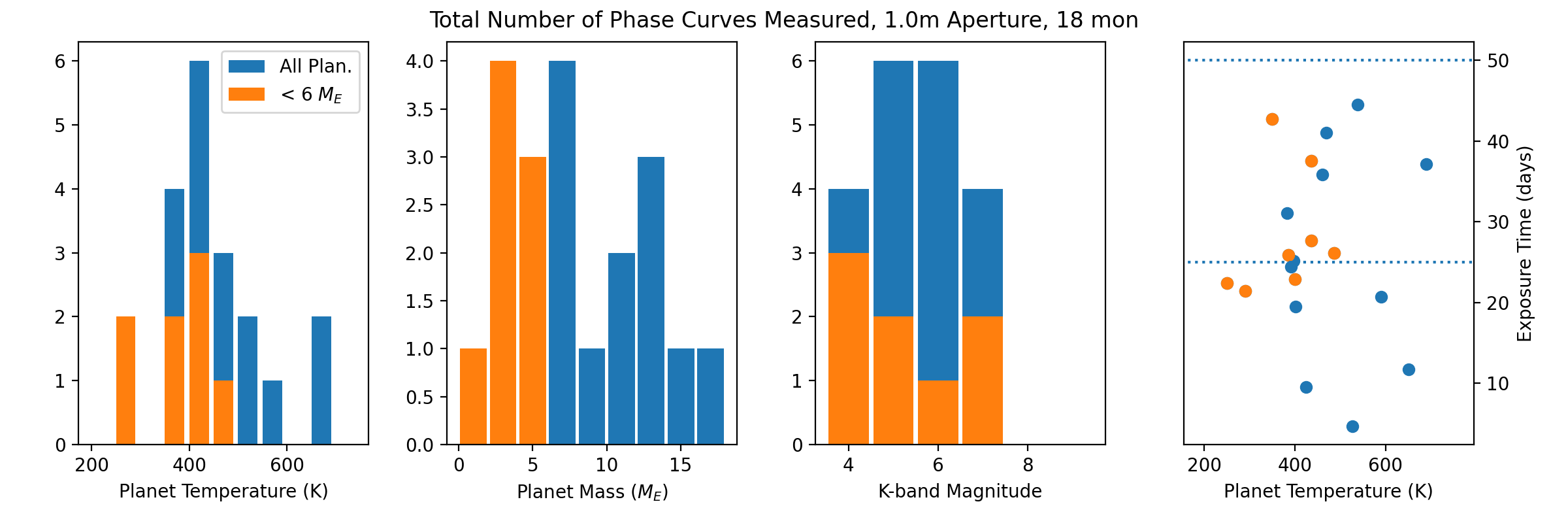}
\includegraphics[width=0.95\textwidth]{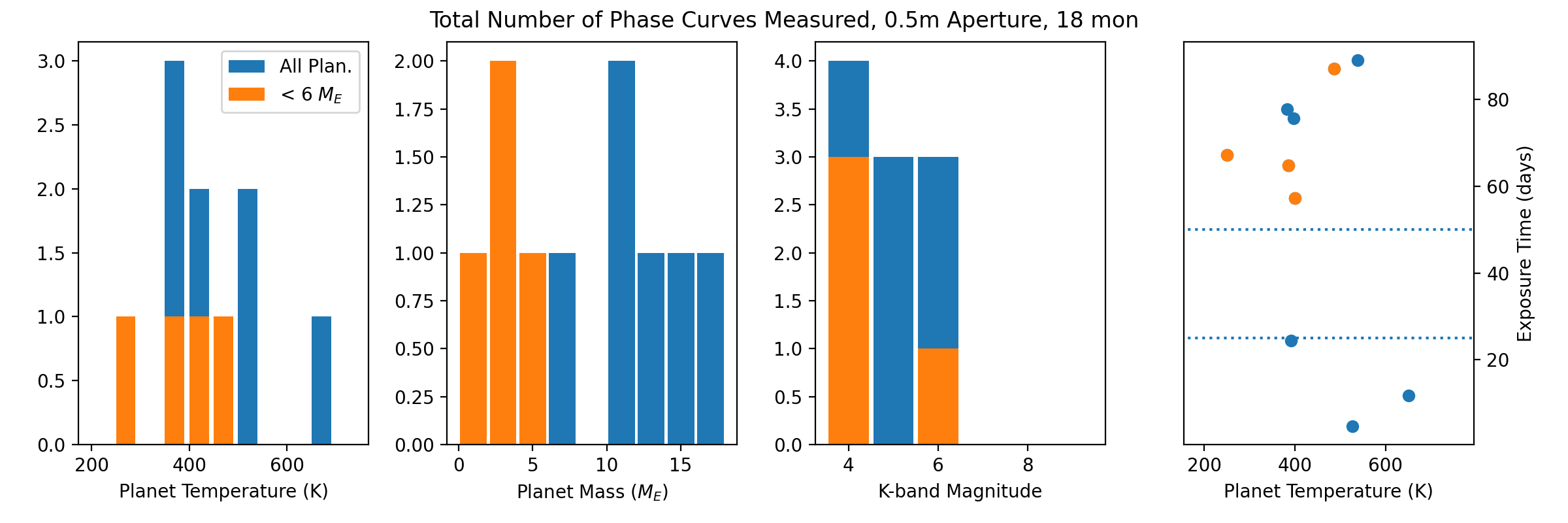}
\caption{Distribution of planet equilibrium temperatures, minimum masses, K-band magnitudes of the host stars, and total exposure times for a survey of phase curve measurements.  The orange bars denote potentially rocky planets ($M_{p} < 6 M_{E}$).  Top: 2-meter aperture, middle: 1-m, bottom: 0.5-m}
\label{fig:phasecurve}
\end{figure*}

Our results are summarized in Figure~\ref{fig:phasecurve}. In total, we determined that phase curves for 24 planets could be measured to the same precision as PCb in 18 months with a 2-meter telescope; this number drops to 20 planets for a 1-m telescope and 10 planets for a 0.5-m telescope.  There is a similar distribution of planet properties, but the exposure time per planet is naturally much longer.  In particular, for the 0.5-m telescope, only 3 Neptune-class planets can be observed with exposure times of less than 25 days, while the 1-m and 2-m telescopes have 3-4 terrestrial planets that could be measured with total exposure times less than 25 days.  

\begin{figure*}[t]
\centering
\includegraphics[width=0.95\textwidth]{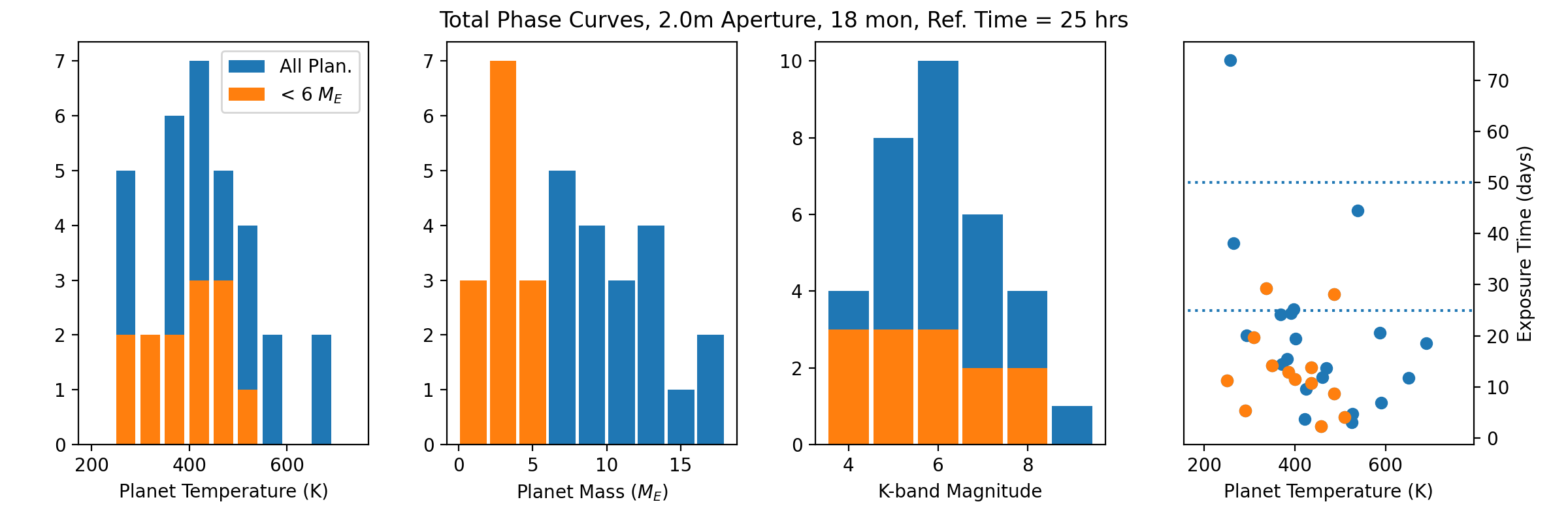}
\caption{Same as the top of Figure~\ref{fig:phasecurve} but with a reference exposure time of 25 hours instead of 100 hours, matching the S/N per target of a 1-m telescope. The number of planets observed increases significantly, demonstrating that the 18-month total exposure time limit is the limited factor in this scenario.}
\label{fig:25hrs}
\end{figure*}

We also examined how the numbers change if the required exposure time for the 2-meter aperture is decreased by 4$\times$, to achieve the same precision as a 1-m aperture (Figure~\ref{fig:25hrs}).  We find that for phase curves, the numbers increase significantly; the 18-month total exposure time limit was the limiting factor, and now more planets can be observed.  In this scenario, we could observe phase curves for 33 planets in 18 months, and most of them would have total exposure times less than 30 days.

As stated above, these yield numbers are only a rough estimate of the actual potential mission yield.  In particular, we excluded targets with more than one planet in the system with a similar thermal flux signal; in principle, this requirement could be relaxed by utilizing sophisticated fitting routines to determine the contribution from each planet.  However, we also used a estimate for the zodiacal dust contribution based on the value for PCb, which is at a high ecliptic latitude; this will increase somewhat for low ecliptic latitudes, though we can design the observing schedule to minimize the contribution in the line of sight by pointing directly away from the Sun.  Additional work on including individual target properties is necessary to determine an exact target list and further improve science yield estimates.

\section{Technology Requirements and Readiness}
\label{sec:trl}

As we demonstrate above, the practical limit on the measurement precision achievable with the PIE technique for bright nearby stars is dependent on the total simultaneous wavelength coverage, the contribution from astrophysical and instrument noise sources at long wavelengths, and the absolute noise floor achievable after integrating over the full exposure time.  For a specific wavelength and noise performance, the overall mission science yield is then dependent primarily on the aperture size and overall throughput, since the total collecting area also affects the amount of astrophysical background and therefore the background-limiting target magnitude. With these factors in mind, we now address the readiness of the telescope and instrument technologies best positioned to achieve the MIRECLE performance requirements.

\subsection{Telescope} 
\label{sec:tel}

The overall MIRECLE observatory would require a cooled telescope and optics, in order to limit the thermal background in the MIR.  However, since the telescope serves only as a point-source light bucket with no need to image the unresolved star or surrounding FOV and is limited to IR wavelengths, relatively relaxed requirements for the primary mirror performance are acceptable. The primary driving requirements on the observatory would be a relatively cold primary mirror and optics ($T_{obs} \leq 35$K) 

To meet the observatory temperature requirement, a combination of radiative and mechanical cooling could be employed, similar to the Spitzer and JWST architectures. MIRECLE would most likely need to operate in either an Earth-trailing orbit like Spitzer, an orbit at the Sun-Earth L2 similar to JWST, or some other orbital configuration that maximizes time away from the Earth's thermal radiation. Using a telescope barrel and sunshields, Spitzer's inner sunshield surface was able to reach 28K \citep{SpitzerHandbook2021}; assuming the proper orbit and pointing strategy, optimized radiative cooling designs could be expected to achieve $30-35$K as an equilibrium temperature (i.e. equilibrium between energy radiated versus parasitic heating due to the spacecraft and stray radiation from either the Sun, Earth, or Moon). 

\subsection{Instrument} 
\label{sec:inst}

The MIRECLE instrument would be a spectrometer using fixed dispersive elements (prism, grating and/or grism) to capture the full wavelength range (1 - 18~{\um} maximum, 2.5 - 16.7~{\um} minimum) simultaneously, and a focal-plane array and electronics that can achieve a flux measurement stability/repeatability of $\leq 5$ ppm over 100 - 600 hours after post-processing to remove known systematics.  The wavelength and flux measurement stability requirements on the detector array would be the most challenging aspect, and will most likely only be possible through a combination of a stable and well-characterized detector array and electronics, and a real-time calibration system that continually provides a stable illumination to re-calibrate the detector gain and offset over the length of the observation and allow for highly effective removal of systematic trends.

These instrument and detector characteristics can potentially be met using existing technologies. \citet{Roellig2020} examined the readiness and development paths of various detector technologies for a similar science case for the Origins Space Telescope, focusing on three technology options:  Si:As impurity band conduction (IBC) arrays, HgCdTe arrays, and transition edge superconductor (TES) bolometer arrays. A laboratory experiment that incorporates the types of instrument control and calibration features described above has been built and is currently being tested at NASA/Goddard Space Flight Center \citep{Staguhn2019}. In Sections \ref{sec:ibc} - \ref{sec:tes} we discuss potential detector technologies, their current demonstrated performance and availability, and the obstacles that need to be addressed to meet MIRECLE requirements, and we examine the design considerations for a continuous calibration system in Section~\ref{sec:cal}. 

\subsubsection{Impurity Band Conduction Detectors} 
\label{sec:ibc}
Si:As impurity band conduction (IBC) detectors have served as a core detector technology for a number of space-based MIR instruments. In particular, the Spitzer IRAC instrument utilized Si:As detectors to achieve ground-breaking space-based NIR and MIR photometric exoplanet measurements, and the JWST/MIRI instrument uses similar arrays.  The IRAC detectors achieved excellent sensitivity and stability of $\sim$35 ppm \citep[e.g.][]{Tamburo2018}, and the MIRI detectors are expected to achieve similar or moderately improved stability \citep{Bright2016}. Upcoming transit and phase curve observations with JWST/MIRI will push the limits of the noise performance, and on-orbit calibration and post-processing are expected to measure and mitigate the instrument instabilities to some level and reveal what can be corrected in the presence of drifts in the instrument and in the astrophysical sources \citep{Greene2016}.

However, limiting factors in stability for IBC detectors are currently not sufficiently understood to determine whether they could achieve 5 ppm over long-duration measurements. The intrinsic stability of existing Si:As detector systems may be driven by fundamental detector materials properties, the current cryogenic detector readout circuitry (which is not optimized for operation at cryogenic temperatures), or other instabilities in the system. Therefore, a detailed study of instabilities in MIR detectors would be needed to determine where technology investments will be most effective.

The main obstacle for the future use of IBC detectors will be the commercial availability of these detectors. At this time, commercial suppliers for these detectors that previously fabricated the arrays for such missions as Spitzer, WISE, and JWST (Raytheon and DRS) have discontinued fabrication efforts in this wavelength range. This is also a result of lack of need of these technologies for applications for the Department of Defense. While large contracts could be used to re-start the development and production of these detectors, this would be associated with significant cost. 

\subsubsection{HgCdTe Detectors}
HgCdTe detector array technology has matured dramatically over the years, and high-performance devices with a cutoff wavelength of 5 \um have been used for JWST and WISE, while the Roman Space Telescope will be using 2.5 \um cutoff HgCdTe devices. With large-format arrays, high well-depths, and relatively high operating temperatures, these devices have become the workhorse of space-based NIR astronomy.  However, the performance of longer cutoff devices had been limited by larger tunneling dark currents, and the softer material of these arrays (due to increased mercury content) can lead to an increased likelihood of forming defects/dislocations that contribute to trap-assisted dark currents \citep{Carmody2003}.

Longer-wavelength HgCdTe arrays are now being developed by University of Rochester (UR) infrared detector group and Teledyne Imaging Systems (TIS) in support of the NEO Surveyor mission concept, with long-wavelength cut-offs of 10.5 \citep{McMurtry2016}, 13 \citep{Cabrera2019}, and now 15 \um \citep{Cabrera2020} now being demonstrated. These devices employed a proprietary design to mitigate quantum tunneling dark currents, and yielded three devices with excellent read noise ($\sim 30 e^-$), QE $\ge 80\%$ in the 6 - 12~{\um} wavelength range (without anti-reflective coating), and reduced expected tunneling dark current behavior. One of those three devices tested by \citet{Cabrera2020} has a cutoff wavelength of 16.7~{\um} at a temperature of 30 K and is bonded to a HAWAII-1RG multiplexer (MUX). The characterization of this device showed that the performance was limited by tunneling dark currents when a modest amount of reverse bias is applied to increase the pixel well depth. These results are very encouraging since the tunneling dark currents showed an improvement in the TIS proprietary design of this device by several orders of magnitude, and motivated our long-wavelength sensitivity analysis in \S\ref{sec:retrieval}.

Additionally, this was the first attempt in producing these devices; with further enhancement to the device design, the tunneling currents may be conceivably mitigated further. Another option for avoiding tunneling dark currents for future devices would be to bond these devices to a MUX based on a capacitive trans-impedance amplifier (CTIA) structure that can provide large well depths while operating at lower biases where the contribution from tunneling currents are minimal. The HAWAII MUX was chosen by UR for the 15 {\um} devices due to their lower power dissipation relative to CTIA structure MUXes. Temperature-controlled operation at 25 K would also minimize the dark current in the current devices, while maintaining the highest level of operability. Space qualified readout of these detectors will be straight-forward, since the electronics developed for JWST would be sufficient and provide a high-TRL entry level point.

\subsubsection{Transition Edge Sensors (TES)}
\label{sec:tes}
TES detectors have been measured to produce white, flat, and featureless noise spectra down to $\sim$10 mHz or below. Compared with semiconducting bolometers or kinetic inductance detectors (KIDs), TES have demonstrated superior stability on timescales of several minutes. Even when 1/f noise is detectable, it is attributed to the SQUID MUX readout rather than the TES device itself, and therefore can be removed as a common mode feature (since each SQUID MUX reads out several detectors). \citet{Staguhn2019} examined the stability of TES arrays in the laboratory, showing that the power spectrum of their detectors shows a signal between 1 Hz and several tens of seconds that is almost flat, while at times greater than $\sim$ 1 minute there is a slight upturn in the noise. However, the increase in noise is moderate, in particular significantly shallower than 1/f, i.e. the noise increase vs time is more modest than it would be in the 1/f case.

The physics of TES detectors themselves is well understood, so the performance of those devices can be reliably designed. TESs are also flexible in terms of readout technologies: The devices can directly be read out with relatively low bandwidth (a few MHz) time domain digital warm electronics. Achievable multiplexing factors for time domain readout systems would not be an issue, since the number of TES pixels required for a MIR spectrometer as described here is modest ($<1000$). Alternatively, microwave multiplexers have been developed at NIST Boulder, which could be used for GHZ frequency domain readout systems that are used for other superconducting detector technologies such as MKIDs.

The most significant instrument design challenge for TES and KID detectors is the required sub-Kelvin operating temperatures necessary to reach optimal sensitivity. For the expected temperatures of $\sim$100 mK, inclusion of adiabatic demagnetization refrigeration (ADR) along with the required cryogenic cooling to reach the ADR operating temperature of approximately 4K is necessary.  Several mission concepts, most notably the Hitomi mission \citep{Shirron2018}, have developed this technology to high TRL, but the requirements on the instrument design are substantial.

\subsubsection{Continuous Calibration}
\label{sec:cal}

For dimmer targets or phase-curve measurements requiring a stability of many hours to reach the required shot-noise uncertainty, the need for continuous calibration of the detector response becomes essential. \citet{Staguhn2019} describes such a calibration system that uses a black body source as a reference signal that is simultaneously observed with the exoplanet by the detectors in the spectrometer. A similar calibrator has been used on the JWST/MIRI instrument for instrument testing \citep{Wright2015}, but is not designed for use in flight. 

By turning the tungsten filament on and off (or otherwise modulating the signal) the gain of the detector can effectively be measured at each on/off transition; modulating on the order of once per minute is sufficient to monitor the gain. There is no penalty on the achieved sensitivity of the measurement,  since the black body signal power is chosen to be sufficiently small as not to increase the photon noise budget materially due to the significant photon noise from the bright stellar hosts. Implementation of this calibration scheme will be an important step in providing the required long term stability of the spectrometer used in MIRECLE.

\section{\revision{Mission-Level Science Impact}}
\label{sec:discussion}

We have demonstrated that with the appropriate simultaneous wavelength coverage and measurement precision, a modestly-sized space telescope could detect and analyze the atmospheres of non-transiting planets around the nearest M-stars, including Proxima Cen b.  We have also reviewed the technological readiness of the instrumentation needed to achieve the required performance, identifying several potential detector technologies that may be able to meet requirements.  Below we discuss the overall potential science returns from the MIRECLE survey science described above, \revision{as well as additional science beyond the PIE survey of planets with $M~\text{sin}~i < 20$ that could be accomplished with MIRECLE. Finally, we address the question of complementarity between the capabilities of MIRECLE and those of upcoming exoplanet-related space-based observatories}.

\subsection{\revision{Primary} Survey Science Yield}
\label{sec:survey}

One of the primary science questions that MIRECLE would answer is the distribution of a rocky planets having tenuous, clear, or cloudy atmospheres. Planets orbiting in the habitable zones (HZs) of M dwarf stars have relatively short orbital periods (5 -- 40 days) and receive higher levels of UV flux than planets in the HZs of G stars.  Extreme UV flux can enhance atmospheric escape to the point of yielding only a tenuous atmosphere.  M dwarfs tend to be most active shortly after formation ($<$ 1 Gyr), thus potentially impacting planets that formed in-situ.  Planets that have migrated into the HZ after 1 Gyr are more likely to retain appreciable atmospheres.  Additionally, the secondary atmospheres of impacted (e.g., in-situ formation) planets could be replenished from infalling cometary material. Understanding what fraction of temperate terrestrial planets with short orbital periods have substantial atmospheres can provide information on the formation and migration of these systems and place our Earth into a broader context.

Similarly, a PIE survey of planets from 1.3 - 20 Earth masses would directly explore the super-Earth/sub-Neptune boundary. At orbital periods less than 50 days, planets below 1.6 Earth radii are thought to be predominantly rocky \citep{Rogers2015}, but it’s not known if they formed that size or were once larger and subsequently lost their primordial atmospheres.  The Kepler observatory has shown us that there is a gap in the occurrence rate of planets between 1.5 and 2.0 Earth radii \citep[e.g.,][]{Fulton2017,Petigura2022}.  This paucity of planets supports the idea that a sub-population of close-in sub-Neptune planets undergo a relatively quick process of atmospheric mass loss that leaves them with a rocky core measuring up to 1.6 Earth radii and potentially a thin, high-mean-molecular-weight atmosphere (i.e., super Earths).  A large characterization survey of super-Earth and sub-Neptune size planets would confirm theoretical predictions of the formation and evolution of super-Earths and sub-Neptunes.

\subsection{Additional Potential MIRECLE Science}
\label{sec:moreScience}

Beyond studies of rocky and Neptune-like worlds using PIE, MIRECLE would also have the capability to bridge the gap between the Hot Jupiters being studied with current and near-term missions and the giant planets in our own Solar System. For giant planets with hydrogen-dominated atmospheres ($\gtrsim10$ Earth masses), missions such as JWST and ARIEL will mostly target warm planets ($>600$ K) in thermal emission. This will leave a gap in our understanding between these objects and the solar system giant planets ($<200$ K). Mid-IR spectroscopic phase curves of Jupiter and Saturn exoplanetary analogs would enable transformational science in the chemistry and structure of temperate giant planets around K and M. In these cooler atmospheres, the abundances of \ce{CH4} and \ce{NH3} probe metallicity, non-equilibrium chemistry, and the strength of vertical mixing \citep{Zahnle2014}.  MIRECLE could probe the rich photochemistry expected for these atmospheres, via production of HCN, \ce{C2H2}, and \ce{C2H4} from methane, as is seen in Jupiter beyond 5 {\um}. 

MIRECLE could also advance our understanding of the physics and chemistry of clouds and hazes, which are currently a major source of uncertainty in models of exoplanet atmospheres.  Since most clouds appear only as Rayleigh or gray scatterers in the optical and near-IR, we have poor constraints on the composition of clouds that are inferred in hot Jupiter atmospheres.  A number of cloud species expected to impact the spectra of these planets have known features in the mid-IR \citep[e.g.,][]{Wakeford2015}.  Examples include: \ce{Al2O3} (12 {\um}), \ce{CaTiO3} (13 {\um}, 22 {\um}), and Mg$_X$SiO$_Y$ (8 -- 22 {\um}).  The population of giant exoplanets that could be probed in thermal emission (down to $\sim$250 K) will be cool enough to, for the first time, serve as a bridge between the extrasolar planet population and our own solar system, and enable the unique identification of aerosols that impact the transmission spectra of hot exoplanets as well.

MIRECLE would also enable high-precision MIR time series spectroscopy of brown dwarf atmospheres. Brown dwarfs are self-luminous bodies that form like stars, but are more similar in temperature to warm and hot Jupiters.  As a result, most brown dwarfs have atmospheres that are near-solar in composition; however, there are numerous examples of brown dwarfs with metal-rich and metal-poor atmospheres (e.g., inner galactic disk and halo galaxies, respectively). Unlike hot Jupiters, brown dwarfs provide an isolated laboratory to investigate condensate cloud formation, pressure-temperature profiles, chemistry, non-equilibrium processes, and three-dimensional time-dependent weather.  For example, long-baseline, multi-epoch observations with MIRECLE would reveal the source of previously-observed variability (e.g., short-lived clouds vs. long-lived features at different latitudes).

\subsection{MIRECLE Complementarity with Existing and Planned MIR Space Missions}
\label{sec:exist}

There are already two existing or approved space observatories with NIR/MIR capabilities that are expected to dramatically expand the characterization of transiting exoplanets spanning a wide range of stellar and planetary properties, and we consider their complementarity to MIRECLE. The recently launched JWST is expected to be a transformative tool for transiting exoplanet characterization due to the large collecting area (equivalent to a 5.7-meter circular aperture) and a combined wavelength coverage from 0.6 to 28~{\um} across all the instruments. The Atmospheric Remote-sensing Infrared Exoplanet Large-survey (ARIEL) mission, selected by ESA for launch in 2028, will focus on systematically characterizing a sample of transiting exoplanets using simultaneous measurements of visible/NIR photometry (0.5 - 1.1~{\um}) and low-resolution NIR/MIR spectroscopy (1.1 - 7.8~{\um}).  

Both missions will significantly advance our understanding of the diversity of exoplanet atmospheres through transmission, eclipse, and phase curve measurements of transiting planets \citep{Greene2016}.  However, they will be hampered in their ability to study the non-transiting temperate terrestrial planet population around the nearest stars -- primarily because neither telescope will have the simultaneous wavelength coverage across the NIR and MIR necessary to effectively perform PIE measurements for small and cool planets.

JWST includes multiple NIR instruments, and the MIRI instrument includes a low-resolution spectroscopy mode (LRS) that extends from 5 to $\sim$13~{\um} with a spectral resolving power of $R=100$. MIRI also includes a mid-resolution spectroscopy mode (MRS) which has four different bands spanning 5 - 28~{\um} and several resolving power settings, but this mode uses an image slicer design that makes high-precision time series measurements extremely challenging. \citetalias{Lustig-Yaeger2021} demonstrated that combining observations with the NIR instruments and MIRI-LRS would allow for emission spectroscopy measurements of hot Jupiters with the PIE technique. However, the NIR and MIR instruments cannot be used to observe the same FOV simultaneously, requiring the stitching of different data sets together to fully constrain the stellar and planetary emission. While the absolute flux uncertainty expected for JWST measurements is not especially detrimental (as shown by \citetalias{Lustig-Yaeger2021}), the asynchronous nature of the measurements would introduce a number of challenges for extracting small-amplitude signals, and we do not examine this option further. Alternatively, using only the MIRI-LRS instrument alone cannot achieve sufficient constraints on the MIR flux of PCb due to the inability to constrain the stellar contribution to the measured MIR flux (as we show in Figure~\ref{fig:retrieval_wavelength}).  

Similar restrictions in simultaneous wavelength coverage limit the capabilities for PIE measurements with the ARIEL mission. ARIEL includes a NIR/MIR spectrograph (AIRS) with R=30--200 as well as simultaneous photometry at shorter wavelengths.  However, with only a few photometric measurements at short wavelengths and a spectrograph that only extends out to 8~{\um}, ARIEL would be similarly unable to measure thermal emission for non-transiting temperate planets.

\section{Future Work}
\label{sec:sar}

Thus far, the PIE technique has shown considerable promise when it comes to characterizing the atmospheres of rocky worlds orbiting nearby M-dwarf stars, such as Proxima Cen b. However, untangling the diverse array of complicating astrophysical and instrumental factors will be critical in ensuring that robust planetary information can be inferred using the PIE technique on non-transiting systems.

Beyond successfully implementing the technique on broadband spectroscopic data of large and hot planets with JWST observations (e.g., HAT-P-26b, WASP-17b, etc.), additional model validation work is also needed to evaluate and improve the robustness of our stellar modeling strategies. The impact of inaccurate spectral opacities and stellar variability on the morphology and temporal variation in the stellar SED is a primary issue that must be addressed, since the stellar flux at MIR wavelengths must be removed accurately without biasing planetary parameters. Multi-wavelength observations spanning X-ray to visible wavelengths have shown the ubiquity of stellar variability and flaring in M-stars \citep[e.g.][]{MacGregor2018}, and the impact of stellar variability on exoplanet transmission spectroscopy has been well documented \citep{Rackham2018, Barclay2021}. 

\revision{However, due to a lack of spaceborne MIR spectroscopic capabilities since the end of the Cold Spitzer mission, no broadband MIR spectra of M-stars that are both high-SNR and free of telluric absorption exist, making the assessment of opacity linelists and the impact of stellar flares impossible.  Fortunately, upcoming observations of M-stars with JWST will help significantly in improving our confidence in stellar models in both the spectral and temporal dimensions. Data from approved exoplanet observations, either for systems with no hot planets or for periods where a planet is eclipsed by the star, can be used for initial model validation and variability studies, and new moderate-resolution observations with the MRS mode of MIRI (5 -- 28 {\um}) could provide high-quality ground truth for validating and improving stellar line lists. Multiple stellar model grids are now available for M-star stellar temperatures, and model grid intercomparison can be used to determine differences and potential areas of future model improvement.} 

Similarly, the contribution from additional known or even unknown planets in multi-planet systems needs to be addressed. In theory, RV measurements should detect and constrain the orbital periods and eccentricities of any planets sufficiently large and close to the parent star to produce an appreciable thermal signal; however, as demonstrated by the recent tentative detection of Proxima Cen d, small planets are hard to detect through RV measurements but can produce a large thermal signal if they are hot enough. While multiple planetary signals may provide an opportunity for a ``joint'' constraint on the combined thermal flux and atmospheric absorption, the goal would be to gain constraints on each planet separately by fitting for the variation in thermal flux over the orbital period of each planet. Uninterrupted simultaneous measurements of the emission from both the star and planet over days and even weeks should provide sufficient leverage to differentiate stellar, planetary, and instrumental flux variations - but this would clearly be impacted by uncertainties in the ephemerides for each planet. More algorithmic development, as well as demonstrations with existing telescopes using well-characterized planetary systems, is necessary to validate these data processing strategies. 

There are several other factors that may lead to large uncertainties in the the final constraints that can be obtained for a particular planet.  The unknown orbital inclination of non-transiting planets means that phase-dependent variations in thermal flux from a planet cannot be directly translated into longitudinal temperature constraints, as is done for transiting planets \citep[e.g.][]{Knutson2007}.  Orbital inclinations for the nearest M-star systems may be measured in the future through direct imaging or astrometry, but there may also be additional ways to constrain inclination of unresolved systems.  The contribution from unknown exozodical dust could also contribute flux at MIR wavelengths, and if the temperature of the dust is similar to that of the planet, it would produce a degeneracy that would be hard to break. However, dynamical modeling of the dust in systems with known planets could put constraints on the potential dust distribution and therefore the likelihood of signal confusion. 

Answering these and other related questions will be the focus of future work as they require increasingly more sophisticated stellar and planetary models.  By leveraging the broad and interdisciplinary expertise of the astrophysics, astrobiology, and instrumentation communities, we can continue advancing the PIE technique as a means to study the vast population of nearby, non-transiting exoplanets and their atmospheres.

\section{Conclusion} \label{sec:concl}

In this paper, we have examined the potential science yield for a stand-alone mission (MIRECLE) to characterize the population of cool, non-transiting terrestrial and sub-Neptune planets around the nearest M-dwarf stars using simultaneous NIR/MIR spectroscopy and the Planetary Infrared Excess (PIE) technique to extract the planet's emission spectrum \citep{Stevenson2020,Lustig-Yaeger2021}. We first simulated observations of the nearest planet to Earth, Proxima Cen b, under the assumption that the planet has an Earth-like atmospheric composition and structure. We explored the impact of the aperture size, simultaneous wavelength range, and absolute noise floor on the ability to constrain the composition and temperature of the atmosphere.  We then used the reference exposure times derived for Prox Cen b and the list of all nearby planets from the Exo.MAST database to calculate a potential planet characterization yield under two separate scenarios: a survey of planets using a time-integrated emission spectrum to constrain the temperature, radius and bulk composition of the planets, and a deeper exploration of the atmosphere and/or surface properties of nearby temperate rocky and sub-Neptune planets using thermal phase curves. 

Our results above clearly indicate that, under standard assumptions for astrophysical and instrument noise contributions, a 2-year mission with a 2-meter aperture and simultaneous wavelength coverage from $1-18\;\mu$m can effectively minimize the planet temperature/radius degeneracy and detect multiple atmospheric absorption features for Proxima Cen b and more than 36 other planets with $M_{p}<20\;M_{E}$ and $T_{eq} < 700$K, including 13 potentially rocky planets with $M_{p}<6\;M_{E}$ and $T_{eq} < 600$K -- as long as a noise floor of 5 ppm or better and total integration times of at least 25 days can be achieved. If phase curves are prioritized instead, 28 planets could be examined, including 9 potentially rocky planets.  Diminishing the aperture size to 0.5 meters decreases the number of available targets and the exposure times, but still allows for thermal emission spectroscopy of 20 planets.  

With the publication of the Astro2020 Decadal Survey Report \citep{Astro2020}, initial preparation is beginning for both the Explorer-class mission proposal opportunities as well as the new Probe-class mission opportunities -- one of which will be a Far-IR Probe that could incorporate the MIRECLE instrumentation and science cases.  Now that JWST is beginning its science operations, it is time to look towards the next generation of exoplanet characterization missions.  We predict that the MIRECLE science case and instrumentation requirements needed for the study of non-transiting planets using the PIE technique will be a key part of these future mission portfolios.
\\

This material is based in part upon work performed as part of the CHAMPs (Consortium on Habitability and Atmospheres of M-dwarf Planets) team, supported by the National Aeronautics and Space Administration (NASA) under Grant No. 80NSSC21K0905 issued through the Interdisciplinary Consortia for Astrobiology Research (ICAR) program.  Funding was also provided through the NASA Goddard SEEC (Sellers Exoplanet Environments Collaboration) team, which is funded in part by the NASA Planetary Science Division’s Internal Scientist Funding Model, and by NASA ROSES grants 17-APRA17-0126 and 18-SAT18-0024. We also thank T. Barclay, D. Fixsen, D. Koll, E. Sharp, and G. Villanueva for valuable conversations and suggestions on the analysis and the manuscript.


\bibliography{main}

\appendix

\section{Target Lists for Integrated Emission Spectroscopy and Phase Curve Spectroscopy} \label{sec:appendix:tables} 

Each of the target lists are organized based on increasing aperture - all the targets that are amenable to measurements with a 0.5-meter aperture would also be included in the targets for a 1-meter and 2-meter apertures. The horizontal lines indicate the breakpoints between 0.5, 1, and 2 meters.

\startlongtable
\begin{deluxetable*}{lrcc|cccc}
\tablewidth{0.98\textwidth}
\tabletypesize{\footnotesize}
\tablecaption{Stellar and planetary parameters for targets amenable to integrated emission spectroscopy with the three different aperture sizes (0.5, 1.0, and 2.0 meters).  \label{tab:targets-em}}
\tablehead{\colhead{Name} & \colhead{Dist} & \colhead{Kmag} & \colhead{T$_{eff}$}  & \colhead{$M~\text{sin}~i$} & \colhead{T$_{eq}$}  & \colhead{Orb. Period}  & \colhead{ESM-15{\um}} \\
           \colhead{} & \colhead{(pc)} & \colhead{} & \colhead{(K)}  & \colhead{(M$_E$)} & \colhead{(K)}  & \colhead{(days)}  & \colhead{} }
\startdata
Proxima Cen b & 1.3 & 4.4 & 3050 & 1.3 & 250 & 11.2 & 46.5 \\
GJ 96 b & 11.9 & 5.6 & 3785 & 19.7 & 258 & 73.9 & 23.8 \\
GJ 687 b & 4.6 & 4.5 & 3413 & 17.2 & 264 & 38.1 & 79.1 \\
Ross 128 b & 3.4 & 5.7 & 3192 & 1.4 & 309 & 9.9 & 22.8 \\
GJ 251 b & 5.6 & 5.3 & 3451 & 4.0 & 349 & 14.2 & 27.5 \\
Gl 686 b & 8.2 & 5.6 & 3656 & 6.6 & 383 & 15.5 & 37.5 \\
GJ 411 b & 5.7 & 3.5 & 3601 & 2.7 & 386 & 12.9 & 43.3 \\
HD 102365 b & 9.3 & 3.5 & 5630 & 15.9 & 390 & 122.1 & 36.9 \\
GJ 338 B b & 6.3 & 4.1 & 4005 & 10.3 & 392 & 24.4 & 63.2 \\
HD 211970 b & 13.1 & 5.7 & 4127 & 13.0 & 397 & 25.2 & 43.5 \\
GJ 15 A b & 3.6 & 4.0 & 3607 & 3.0 & 400 & 11.4 & 45.5 \\
GJ 720 A b & 15.6 & 6.1 & 3837 & 13.6 & 401 & 19.5 & 44.2 \\
GJ 1265 b & 10.2 & 8.1 & 3236 & 7.4 & 422 & 3.7 & 95.3 \\
GJ 480 b & 14.2 & 6.7 & 3381 & 13.2 & 424 & 9.6 & 61.3 \\
Gl 49 b & 9.8 & 5.4 & 3805 & 5.6 & 436 & 13.9 & 30.9 \\
CD Cet b & 8.6 & 7.8 & 3130 & 4.0 & 458 & 2.3 & 82.1 \\
GJ 536 b & 10.4 & 5.7 & 3685 & 5.4 & 487 & 8.7 & 33.2 \\
GJ 674 b & 4.6 & 4.9 & 3600 & 11.1 & 527 & 4.7 & 278.3 \\
HD 177565 b & 16.9 & 4.5 & 5627 & 15.1 & 539 & 44.5 & 42.1 \\
HD 190007 b & 12.7 & 4.8 & 4610 & 16.5 & 651 & 11.7 & 108.6 \\
\hline
HD 180617 b & 5.9 & 4.7 & 3534 & 12.2 & 202 & 105.9 & 13.4 \\
GJ 229 A c & 5.8 & 4.2 & 3913 & 7.3 & 220 & 122.0 & 11.9 \\
GJ 3323 b & 5.4 & 6.7 & 3159 & 2.0 & 291 & 5.4 & 41.3 \\
GJ 422 b & 12.7 & 7.0 & 3323 & 11.1 & 293 & 20.1 & 26.9 \\
GJ 625 b & 6.5 & 5.8 & 3499 & 2.8 & 336 & 14.6 & 17.3 \\
GJ 685 b & 14.3 & 6.1 & 3816 & 9.0 & 369 & 24.2 & 22.9 \\
LSPM J2116+0234 b & 17.6 & 7.3 & 3475 & 13.3 & 371 & 14.4 & 37.8 \\
LTT 1445 A b & 6.9 & 6.5 & 3337 & 2.2 & 436 & 5.4 & 26.0 \\
GJ 3082 b & 16.6 & 7.0 & 3910 & 8.2 & 460 & 11.9 & 29.7 \\
HD 238090 b & 15.2 & 6.1 & 3933 & 6.9 & 469 & 13.7 & 26.8 \\
GJ 393 b & 7.0 & 5.3 & 3579 & 1.7 & 487 & 7.0 & 15.5 \\
GJ 1151 b & 8.0 & 7.6 & 3114 & 2.5 & 510 & 2.0 & 46.0 \\
GJ 3779 b & 13.7 & 8.0 & 3324 & 8.0 & 526 & 3.0 & 77.0 \\
GJ 1214 b & 14.6 & 8.8 & 3250 & 8.2 & 588 & 1.6 & 120.1 \\
GJ 3942 b & 16.9 & 6.3 & 3867 & 7.1 & 591 & 6.9 & 34.2 \\
HD 154088 b & 18.3 & 4.8 & 5374 & 6.6 & 690 & 18.6 & 23.5 \\
\hline
HIP 38594 b & 17.8 & 6.2 & 4086 & 8.1 & 309 & 60.7 & 8.6
\enddata
\end{deluxetable*}

\startlongtable
\begin{deluxetable*}{lrcc|cccc}
\tablewidth{0.98\textwidth}
\tabletypesize{\footnotesize}
\tablecaption{Stellar and planetary parameters for targets amenable to spectroscopic phase curves with the three different aperture sizes (0.5, 1.0, and 2.0 meters).  \label{tab:targets-pc}}
\tablehead{\colhead{Name} & \colhead{Dist} & \colhead{Kmag} & \colhead{T$_{eff}$}  & \colhead{$M~\text{sin}~i$} & \colhead{T$_{eq}$}  & \colhead{Orb. Period}  & \colhead{ESM-15{\um}} \\
           \colhead{} & \colhead{(pc)} & \colhead{} & \colhead{(K)}  & \colhead{(M$_E$)} & \colhead{(K)}  & \colhead{(days)}  & \colhead{} }
\startdata
Proxima Cen b & 1.3 & 4.4 & 3050 & 1.3 & 250 & 11.2 & 46.5 \\
Gl 686 b & 8.2 & 5.6 & 3656 & 6.6 & 383 & 15.5 & 37.5 \\
GJ 411 b & 5.7 & 3.5 & 3601 & 2.7 & 386 & 12.9 & 43.3 \\
GJ 338 B b & 6.3 & 4.1 & 4005 & 10.3 & 392 & 24.4 & 63.2 \\
HD 211970 b & 13.1 & 5.7 & 4127 & 13.0 & 397 & 25.2 & 43.5 \\
GJ 15 A b & 3.6 & 4.0 & 3607 & 3.0 & 400 & 11.4 & 45.5 \\
GJ 536 b & 10.4 & 5.7 & 3685 & 5.4 & 487 & 8.7 & 33.2 \\
GJ 674 b & 4.6 & 4.9 & 3600 & 11.1 & 527 & 4.7 & 278.3 \\
HD 177565 b & 16.9 & 4.5 & 5627 & 15.1 & 539 & 44.5 & 42.1 \\
HD 190007 b & 12.7 & 4.8 & 4610 & 16.5 & 651 & 11.7 & 108.6 \\
\hline
GJ 687 b & 4.6 & 4.5 & 3413 & 17.2 & 264 & 38.1 & 79.1 \\
GJ 3323 b & 5.4 & 6.7 & 3159 & 2.0 & 291 & 5.4 & 41.3 \\
GJ 251 b & 5.6 & 5.3 & 3451 & 4.0 & 349 & 14.2 & 27.5 \\
GJ 720 A b & 15.6 & 6.1 & 3837 & 13.6 & 401 & 19.5 & 44.2 \\
GJ 480 b & 14.2 & 6.7 & 3381 & 13.2 & 424 & 9.6 & 61.3 \\
LTT 1445 A b & 6.9 & 6.5 & 3337 & 2.2 & 436 & 5.4 & 26.0 \\
Gl 49 b & 9.8 & 5.4 & 3805 & 5.6 & 436 & 13.9 & 30.9 \\
GJ 3082 b & 16.6 & 7.0 & 3910 & 8.2 & 460 & 11.9 & 29.7 \\
HD 238090 b & 15.2 & 6.1 & 3933 & 6.9 & 469 & 13.7 & 26.8 \\
GJ 3942 b & 16.9 & 6.3 & 3867 & 7.1 & 591 & 6.9 & 34.2 \\
HD 154088 b & 18.3 & 4.8 & 5374 & 6.6 & 690 & 18.6 & 23.5 \\
\hline
GJ 96 b & 11.9 & 5.6 & 3785 & 19.7 & 258 & 73.9 & 23.8 \\
GJ 422 b & 12.7 & 7.0 & 3323 & 11.1 & 293 & 20.1 & 26.9 \\
Ross 128 b & 3.4 & 5.7 & 3192 & 1.4 & 309 & 9.9 & 22.8 \\
GJ 625 b & 6.5 & 5.8 & 3499 & 2.8 & 336 & 14.6 & 17.3 \\
GJ 685 b & 14.3 & 6.1 & 3816 & 9.0 & 369 & 24.2 & 22.9 \\
LSPM J2116+0234 b & 17.6 & 7.3 & 3475 & 13.3 & 371 & 14.4 & 37.8 \\
GJ 1265 b & 10.2 & 8.1 & 3236 & 7.4 & 422 & 3.7 & 95.3 \\
CD Cet b & 8.6 & 7.8 & 3130 & 4.0 & 458 & 2.3 & 82.1 \\
GJ 393 b & 7.0 & 5.3 & 3579 & 1.7 & 487 & 7.0 & 15.5 \\
GJ 1151 b & 8.0 & 7.6 & 3114 & 2.5 & 510 & 2.0 & 46.0 \\
GJ 3779 b & 13.7 & 8.0 & 3324 & 8.0 & 526 & 3.0 & 77.0 \\
GJ 1214 b & 14.6 & 8.8 & 3250 & 8.2 & 588 & 1.6 & 120.1
\enddata
\end{deluxetable*}

\end{document}